\documentclass[journal]{IEEEtran}
\usepackage{tikz}
\usetikzlibrary{arrows,positioning,fit,shapes,calc,decorations.pathreplacing,matrix,patterns}

\usepackage{amsmath}
\usepackage{subfigure}
\usepackage{lineno,hyperref}
\usepackage{xcolor}
\usepackage{extarrows}
\usepackage{algorithm,algorithmicx}
\usepackage[noend]{algpseudocode}
\usepackage{amssymb}
\usepackage{graphicx}
\usepackage[justification=centering]{caption}
\usepackage{cite}
\usepackage{geometry}
\usepackage{tikz}
\usepackage{amsmath}

\usepackage{stfloats}
\usepackage{amsthm}

\newcommand{\ba}{\boldsymbol{a}}

\newcommand{\bb}{\boldsymbol{b}}

\newcommand{\bx}{\boldsymbol{x}}

\newcommand{\bp}{\boldsymbol{p}}
\newcommand{\bs}{\boldsymbol{s}}
\newcommand{\bq}{\boldsymbol{q}}
\newcommand{\br}{\boldsymbol{r}}

\newcommand{\bX}{\boldsymbol{X}}

\newcommand{\bP}{\boldsymbol{P}}
\newcommand{\bY}{\boldsymbol{Y}}
\newcommand{\bA}{\boldsymbol{A}}

\newcommand{\bR}{\boldsymbol{R}}

\newcommand{\bW}{\boldsymbol{W}}
\newcommand{\bI}{\boldsymbol{I}}
\newcommand{\bF}{\boldsymbol{F}}
\newcommand{\bC}{\boldsymbol{C}}
\newcommand{\bD}{\boldsymbol{D}}
\newcommand{\bU}{\boldsymbol{U}}
\newcommand{\bV}{\boldsymbol{V}}
\newcommand{\bQ}{\boldsymbol{Q}}
\newcommand{\bH}{\boldsymbol{H}}
\newcommand{\bS}{\boldsymbol{S}}
\newcommand{\bG}{\boldsymbol{G}}

\newcommand{\bXi}{\boldsymbol{\Xi}}

\newcommand{\bLambda}{\boldsymbol{\Lambda}}

\newcommand{\N}{\mathcal{N}}
\newcommand{\mD}{\mathcal{D}}

\newcommand{\mA}{\mathcal{A}}

\newcommand{\bnu}{\boldsymbol{\nu}}

\newcommand{\Tr}{\text{Tr}}

\newcommand{\herm}{^{\textrm{H}}}
\newcommand{\tra}{^{\textrm{T}}}

\newcommand{\CN}{\mathcal{CN}}
\newcommand{\MN}{\mathcal{MN}}

\newcommand{\bPhi}{\boldsymbol{\Phi}}

\usetikzlibrary{arrows,positioning,fit,shapes,calc,decorations.pathreplacing,matrix,patterns}

\newgeometry{left=1.5cm, right=1.5cm, top=1.5cm, bottom=1cm}
\tikzstyle{factornode} = [draw, fill=white, circle, inner sep=1pt, minimum size=1.4cm]
\tikzstyle{funnode} = [draw, rectangle,fill=black!100, minimum size = 8mm]

\begin{document}
\title {{Matrix Factorization Based Blind Bayesian Receiver for Grant-Free Random Access in mmWave MIMO mMTC}}
\author{Zhengdao~Yuan, Fei Liu, Qinghua Guo, \IEEEmembership{Senior Member, IEEE},
Xiaojun Yuan, \IEEEmembership{Senior Member, IEEE}, Zhongyong Wang, and Yonghui Li, \IEEEmembership{Fellow, IEEE}  
	\thanks{Z. Yuan is with the Artificial Intelligence Technology Engineering Research Center, Open University of Henan, Zhengzhou 450002, China. He was with the School of Electrical, Computer and Telecommunications Engineering, University of Wollongong, Wollongong, NSW 2522, Australia (e-mail: yuan\_zhengdao@foxmail.com).}
	\thanks{F. Liu and Z. Wang are with the School of Geoscience and Techonology and the School of  Electrical and Information Engineering, Zhengzhou University, Zhengzhou 450000, China, (e-mail: ieliufei@hotmail.com, zywangzzu@gmail.com)}
	\thanks{Q. Guo is with the School of Electrical, Computer and Telecommunications Engineering, University of Wollongong, Wollongong, NSW 2522, Australia  (e-mail: qguo@uow.edu.au).}
	\thanks{ X. Yuan is with the National Key Laboratory on Wireless Communications, University of Electronic Science and Technology of China, Chengdu 611731, China (e-mail:
		xjyuan@uestc.edu.cn).}
	\thanks{Yonghui Li is with the School of Electrical and Information Engineering, University of Sydney, Sydney,
		NSW 2006, Australia (e-mail: yonghui.li@ sydney.edu.au)}
}
\markboth{Blind Grant-Free Random Access}
{Shell \MakeLowercase{\textit{et al.}}: Bare Demo of IEEEtran.cls for IEEE Journals}

\maketitle

\begin{abstract}

Grant-free random access is promising for massive connectivity with sporadic transmissions in massive machine type communications (mMTC), where the hand-shaking between the access point (AP) and users is skipped, leading to high access efficiency. In grant-free random access, the AP needs to identify the active users and perform channel estimation and signal detection. Conventionally, pilot signals are required for the AP to achieve user activity detection and channel estimation before active user signal detection, which may still result in substantial overhead and latency. In this paper, to further reduce the overhead and latency, we explore the problem of grant-free random access without the use of pilot signals in a millimeter wave (mmWave) multiple input and multiple output (MIMO) system, where the AP performs blind joint user activity detection, channel estimation and signal detection (UACESD). We show that the blind joint UACESD can be formulated as a constrained composite matrix factorization problem, which can be solved by exploiting the structures of the channel matrix and signal matrix. Leveraging our recently developed unitary approximate message passing based matrix factorization (UAMP-MF) algorithm, we design a message passing based Bayesian algorithm to solve the blind joint UACESD problem. Extensive simulation results demonstrate the effectiveness of the blind grant-free random access scheme.
\end{abstract}

\begin{IEEEkeywords}
Grant-free random access, user activity detection, massive machine type communications, millimeter wave communications, signal detection, approximate message passing, matrix factorization.
\end{IEEEkeywords}

\section{Introduction}


{Massive machine type communications (mMTC) is one of the major application scenarios in the fifth generation (5G) wireless communications, which can be used to accommodate massive connections with sporadic transmissions in internet of things (IoT) systems \cite{Boccardi2014,Tullberg2016}. The conventional grant-based access technique needs a handshaking procedure to exchange the control signaling between the access point (AP) and active users to establish the communication links between them, which may result in excessive overhead, long and uncertain latency. Due to the sporadic short-burst machine-type IoT data traffic, the high overhead and long latency can be unacceptable as the communication becomes inefficient due to the small amount of payload data \cite{Islam2014}. Recently, grant-free random access has attracted much attention, where the handshaking procedure is skipped and users can transmit data straight away at any time slot. 
In the 5G and future generation of wireless communications, the  millimeter-wave (mmWave) frequencies between 30-300 GHz will be employed to alleviate the current spectrum shortage in sub-6GHz bands. In addition, the multiple input and multiple output (MIMO) technology is used to further improve the spectrum efficiency and communication reliability. Hence, in this work, we consider the issue of grant-free random access in a mmWave MIMO systems. }

To achieve grant-free random access, the AP normally needs to perform user activity detection to identify active users, channel estimation to acquire channel state information of active users and multi-user signal detection to detect non-orthogonal signals of the active users. 
{User activity detection can be coupled with channel estimation and/or active user signal detection, e.g., pilot assisted joint user activity detection and channel estimation is followed by multi-user signal detection in\cite{Chen2018Sparse, Fu2019Active, Takanori2022LWC, Zhang2020TVT, Zhang2022TWC, Chen2022TWC, Zhang2020IOT}; channel estimation is followed by performing user activity detection and signal detection jointly in \cite{Wei2017Approx} and \cite{Chi2017Message}; and joint user activity detection, channel estimation and active user signal detection are performed with the aid of pilot signals in \cite{Wei2019Message} and \cite{Du2018Joint}. Exploiting the fact that only a small fraction of users in the network are active at a time, joint user activity detection and active user signal detection or joint user activity detection and channel estimation can be formulated as compressive sensing (CS) problems \cite{Liu2018Sparse}, \cite{Zhang2018TVT}, which can be solved using sparse signal recovery algorithms.  In \cite{Wei2017Approx} and \cite{Wang2015Compres}, with the assumption that the channel state information is available at the AP, various methods such as those based on orthogonal matching pursuit (OMP) \cite{Wang2016Dynamic}, approximate message passing (AMP) \cite{Wei2017Approx} and prior-information aided adaptive subspace pursuit \cite{Du2017Eff} were developed for joint user activity detection and active-user signal detection. However, in many scenarios, wireless channels vary over time and have to be estimated frequently. 
In the existing works, pilot signals are often employed for either user activity detection or joint user activity detection and active user signal detection. 
The use of pilot signals can result in a substantial overhead, especially in a MIMO grant-free random access system where the number of channel coefficients to be estimated is relatively large. In addition, the transmission of pilot signals (followed by data) can also lead to considerable latency. To make the grant-free random access more efficient, it is significant to further reduce the overhead and latency, e.g., by reducing the number of pilot signals or even avoiding the use of pilot signals \cite{Zhang2020TVT, Ding2019TWC, Zhang2018TCOMM}}. 

{In this work, we consider the problem of grant-free random access in a mmWave MIMO system as illustrated in Fig. 1, where the AP is equipped with an antenna array and provides service to a number of IoT users in the area. In particular, we focus on uplink transmission. The users do not transmit pilot signals, thereby the overhead and latency due to pilot signals can be avoided, leading to significant improvement in access efficiency. 
However, this leads to a challenging task for the AP, i.e., the AP needs to perform blind joint user activity detection, channel estimation and signal detection (UACESD). In this paper, we show that the blind joint UACESD can be formulated as a constrained composite matrix factorization problem, which can be solved by exploiting the strong constraints (structures) of the channel matrix and signal matrix, i.e., the mmWave MIMO channel matrix exhibits sparsity in the beam domain \cite{Guo2017Mill}, \cite{Xiao2017} and the elements of the signal matrix are discrete-valued, which are randomly drawn from the symbol alphabet. With these constraints, we recover the channel matrix and signal matrix, and determine the rank of the matrices simultaneously, thereby achieving joint blind UACESD. It is worth mentioning that the blind joint UACESD problem is solved by exploiting the constraints of the relevant matrices rather than relying on pilot signals. Then we design a Bayesian method for the formulated constrained composite matrix factorization problem. In particular, leveraging a recently proposed unitary approximate message passing for matrix factorization (UAMP-MF) algorithm \cite{Yuan2022UAMPMF}, a message passing algorithm is designed in this work. Extensive simulations are carried out to demonstrate the effectiveness of the proposed grant-free random access scheme without using pilot signals.}

\begin{figure}[!t]
	\centering
	\includegraphics[width=0.5\textwidth]{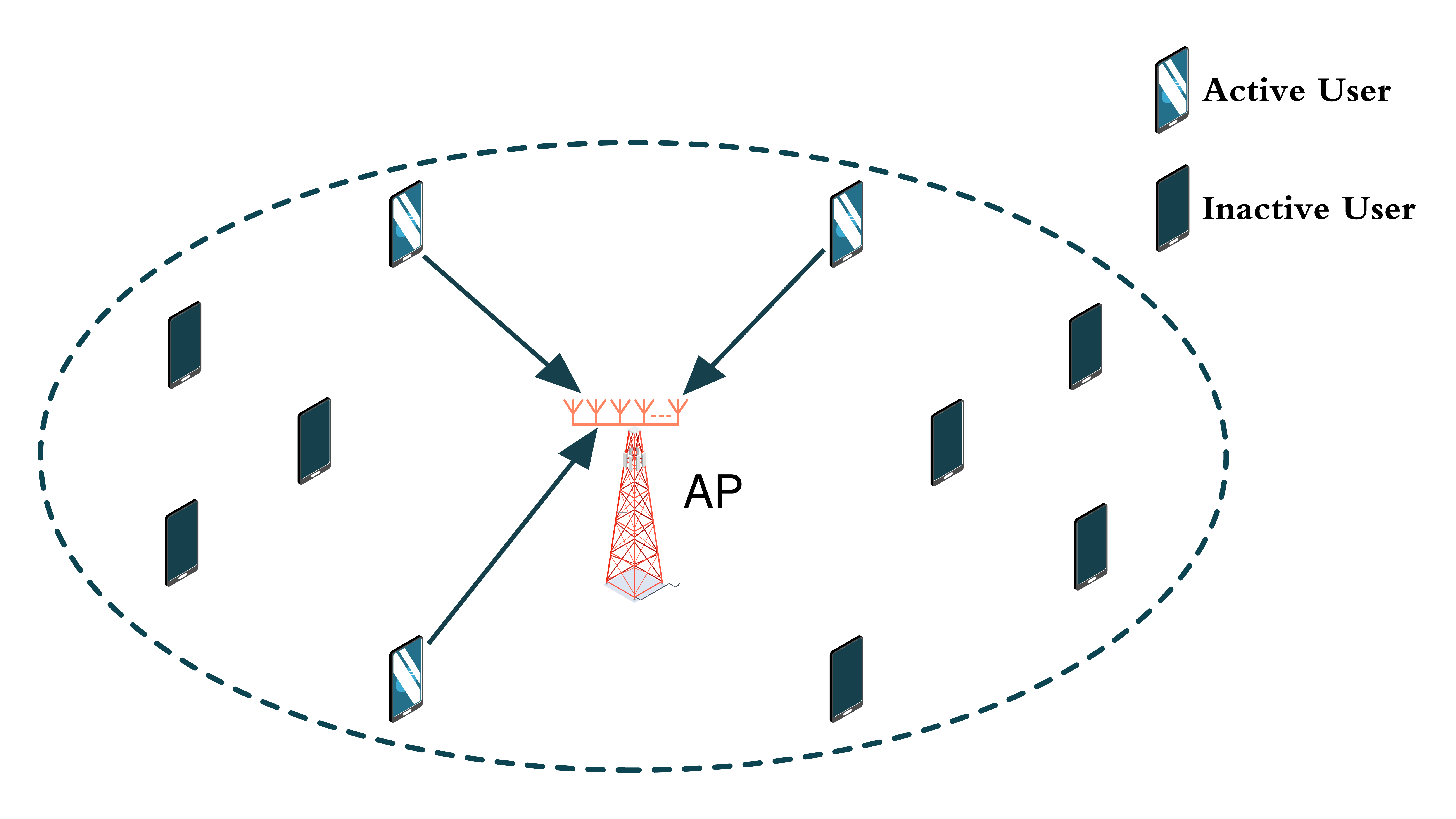}
	\centering
	\caption{Illustration of grant-free random access.}
	\label{fig:RC}
\end{figure}

{The rest of the paper is organized as follows. An brief introduction to (unitary) approximate message passing and its application to matrix factorization are presented in Section II. The system model for grant free random access in mmWave MIMO and the problem formulation of joint blind UACESD are described in Section III. The message passing based Bayesian blind joint UACESD algorithm is developed in Section IV. Numerical simulation results are provided in Section V, followed
by conclusions in Section VI.}

{\textit{Notations}} - The notations used in the paper are as follows. Boldface lower-case and upper-case letters denote vectors and matrices, respectively, and superscript $(\cdot)^T$ represents the transpose operation.
A Gaussian distribution of $x$ with mean {$\hat x$} and variance $\nu_x$ is represented by $\N(x;{\hat x},\nu_x)$. We also simply use $\N(m, v)$ to represent a Gaussian distribution with mean $m$ and variance $v$. Notation $\otimes$ represents the Kronecker product. The relation $f(x)=cg(x)$ for some positive constant $c$ is written as $f(x)\propto g(x)$. We use $\ba\cdot\bb$ and $\ba\cdot/\bb$ to represent the element-wise product and division between vectors $\ba$ and $\bb$, respectively. The notation $\ba^{.-1}$ denotes the element-wise inverse operation to vector $\ba$.
We use $|\bA|]{^{.2}}$ to denote element-wise magnitude squared operation for  $\bA$, and use $||\ba||^2$ to denote the squared $l_2$ norm of $\ba$.  The notation $<\ba>$ denotes the average operation for $\ba$, i.e., the sum of the elements of $\ba$ divided by the number of its elements.The probability density function for the normal random matrix $\bX$ is denoted as $\bX\sim\MN\big(\bX;\hat\bX,\bU_X,\bV_X\big)$, where $\hat\bX$ denotes the mean of the random matrix $\bX$,  $\bU_X$ and $\bV_X$ determine the covariance among rows and columns of $\bX$, respectively.  We use $\textbf{1}$ and  $\textbf{0}$ to denote an all-one vector and an all-zero vector with proper length, respectively. Sometimes, we use a subscript $n$ for $\textbf{1}$, i.e., $\textbf{1}_n$ to indicate its length $n$. The superscript of $\ba^t$ denotes the $t$-th iteration for $\ba$. We use $[\ba]_n$ to denoted the $n$-the element of $\ba$.

\section {Preliminary: {(Unitary) Approximate Message Passing for Matrix Factorization}}

{The AMP algorithm was original developed based on the loopy belief propagation with Gaussian and Taylor-series approximations \cite{Donoho2010message} for compressive sensing with the following model
\begin{equation}
		\bold{y}=\bold{A} \bx+\bold{w},
	\label{y=ax}
\end{equation}
where $\bold{y}$ is an observation vector, $\bold{A}$ is a known measurement matrix, $\bold{x}$ is a (sparse) vector to be recovered and $\bold{w}$ is a noise vector with zero mean and covariance matrix $\beta^{-1} \mathbf{I}$. AMP was later extended to solve estimation problems with a generalized linear observation model in \cite{rangan2011}. AMP enjoys low complexity and it works well in the case of large independent and identically distributed (sub)Gaussian $\mathbf{A}$, but it can easily diverge in the case of a generic matrix $\mathbf{A}$ \cite{Rangan2019convergence}. 

Inspired by the work in \cite{guo2013}, the work in \cite{guo2015approximate} shows that the robustness of AMP can be improved remarkably through simple pre-processing, i.e., performing a unitary transformation to the original linear model \cite{guo2015approximate}.
As any matrix $\mathbf{A}$ has an SVD  $\mathbf{A= U \Lambda V}$ with $\mathbf{U}$ and $\mathbf{V}$ being two unitary matrices, performing a unitary transformation with $\mathbf{U}^H$ leads to the following transformed model
\begin{equation}
	\br=\mathbf{\Phi} \bx+\omega,
	\label{r=uy}
\end{equation}
where $\mathbf{r=U}^H \mathbf{y}$, $\mathbf{\Phi}=\mathbf{U}^H\mathbf{A}=\mathbf{\Lambda}\mathbf{V}$,
$\mathbf{\Lambda}$ is a rectangular diagonal matrix, and ${\omega} = \mathbf{U}^H \mathbf{w}$ remains a zero-mean  Gaussian noise vector with the same covariance matrix  $\beta^{-1} \mathbf{I}$. 
We can apply the vector step size AMP {\cite{rangan2011}} to model \eqref{r=uy}, leading to the first version of UAMP {(called UAMPv1)} shown in Algorithm \ref{UTAMPv1} ({by replacing $\mathbf{r}$ and $\mathbf{\Phi}$ with $\mathbf{y}$ and $\mathbf{A}$ in Algorithm \ref{UTAMPv1} respectively, the original AMP algorithm is recovered})}.
\begin{algorithm}
	\caption{UAMP (UAMPv2 executes operations in [ ])}
	Initialize $\boldsymbol{\tau}_x^{(0)} (\mathrm{or}~{\tau}_x^{(0)})>0$ and ${{\bx}^{(0)} }$. Set $\bs^{(-1)}=\mathbf{ 0 }$ and $t=0$. Define vector ${\boldsymbol{\lambda}=\mathbf{ \Lambda \Lambda}^H \textbf{1}}$. \\
	\textbf{Repeat}
	\begin{algorithmic}[1]
		\State $\boldsymbol{\tau}_p$ = $   \mathbf{|\Phi|}^{.2} \boldsymbol{\tau}^t_x$~~~~~~~~~~~~ $\left[\mathrm{or}~ \boldsymbol{\tau}_p =  \tau^t_x  \boldsymbol{\lambda}\right]$
		\State $ \bp= \mathbf{\Phi}  {{\bx}^{t} } - \boldsymbol{\tau}_{p} \cdot  \bs^{t-1} $
		\State $ \boldsymbol{\tau}_s = \mathbf{1}./ (\boldsymbol{\tau}_p+\beta^{-1} \mathbf{1}) $
		\State  $ \bs^t= \boldsymbol{\tau}_s \cdot (\br-\bp) $
		\State$\mathbf{1}./ \boldsymbol{\tau}_q$ = $ |\mathbf{\Phi}^H |^{.2}  \boldsymbol{\tau}_s$~~~~~~~$\left[\mathrm{or}~ \mathbf{1}./ \boldsymbol{\tau}_q = (\frac{1}{N} \boldsymbol{\lambda}^H \boldsymbol{\tau }_s) \mathbf{1}\right]$
		\State $ \bq = {{\bx}^{t} } + \boldsymbol{\tau}_q \cdot(  \mathbf{\Phi}^H \bs^t)$
		\State $\boldsymbol{\tau}_x^{t+1}$ = $\boldsymbol{\tau}_q \cdot g_{x}' ( \bq, \boldsymbol{\tau}_q)$~~~~~~$\left[\mathrm{or}~ \tau_x^{t+1} \!=\!  \frac{1}{N}  \mathbf{1}^H  \left(\boldsymbol{\tau}_q \cdot g_{x}' ( \bq, \tau_q)\right) \right]$
		\State $ {{\mathbf{x}}^{t+1} } = g_{x}  ( \bq, \boldsymbol{\tau}_q)$
		\State $  t=t+1$
	\end{algorithmic}
	\textbf{Until terminated}
	\label{UTAMPv1}
\end{algorithm}
Applying an average operation to two vectors: $\boldsymbol{\tau}_x$ in Line 7 and $|\mathbf{\Phi}^H |^{.2}  \boldsymbol{\tau}_s$ in Line 5 of {UAMPv1 in} Algorithm \ref{UTAMPv1} leads to the second version of UAMP with lower complexity, where the number of matrix-vector products is reduced from 4 in UAMPv1 (or the original AMP) to 2 per iteration (refer to \cite{Yuan2021} for detailed derivation).

{In the (U)AMP algorithms, the function $g_x(\mathbf{q}, \boldsymbol{\tau}_q )$ is related to the prior of $\mathbf{x}$, which returns a column vector with the $n$th element $[ g_x(\mathbf{q}, \boldsymbol{\tau}_q ) ]_n$ given as 
\begin{equation}
	[g_x(\mathbf{q}, \boldsymbol{\tau}_q ) ]_n
	=
	\frac{\int x_n p(x_n) \mathcal{N} (x_n ; q_n, \tau_{q_n})  d x_n }{\int  p(x_n) \mathcal{N} (x_n ; q_n, \tau_{q_n})  d x_n },
	\label{g_x}
\end{equation}
where $p(x_n)$ represents a prior for $x_n$.
The function $g_x'(\mathbf{q},\boldsymbol{\tau}_q)$ returns a column vector and the $n$th element is denoted by $[ g_x'(\mathbf{q}, \boldsymbol{\tau}_q ) ]_n$, where the derivative is taken with respect to $q_n$.}

In \cite{Yuan2022UAMPMF}, leveraging the variational inference {\cite{jordan1999introduction} and UAMP, an algorithm called UAMP-MF is developed to solve the matrix factorization problem with model
\begin{equation}
	\bR=\bA \bX + \bW
\end{equation}
where $\bR$ is an observation matrix, $\bW$ is a noise matrix, and matrices $\bA$ and  $\bX$ are two factor matrices to be recovered. To achieve high performance, instead of using the mean field approximation
with full factorization, the two matrices are decoupled and are treated as two high dimensional latent variables, leading to the updates of two distributions on the two matrices. By exploiting the structure
of involved variational messages and through a
covariance matrix whitening process, UAMPv1 is incorporated to variational inference to efficiently deal with the distribution updates of the two matrices. The method can be implemented using
message passing with UAMP as its key component. UAMP-MF inherits the low
complexity and robustness of UAMP, and can be used to deal with many matrix factorization problems \cite{Yuan2022UAMPMF}. The UAMP-MF algorithm is shown in Algorithm \ref{alg:UAMPMF}, where the quantity $C$ in Line 25 is computed as
\begin{eqnarray}
	C=\Tr\Big(\big(\bY-\hat\bH\hat\bX\big)\tra\big(\bY-\hat\bH\hat\bX\big)\Big)
	+\Tr\Big(\hat\bX\hat\bX\tra\Tr(\bU_H)\bV_H\nonumber\\
	+\Tr(\bV_X)\bU_X\hat\bH\tra\hat\bH+\Tr(\bV_X)\bU_X\Tr(\bU_H)\bV_H\Big). \label{eq:C}
\end{eqnarray}

In this work, we will formulate the blind joint UACESD as a constrained composite matrix factorization problem, and leveraging the UAMP-MF algorithm, we will develop an efficient message passing algorithm to solve the constrained composite matrix factorization problem.

\begin{algorithm}
	\caption{UAMP-MF}
	\textbf{Initialization}: $\bU_H=\bI_M$, $\bV_H=\bI_N$, $\hat\bH=\textbf{1}_{MN}$, $\bV_X=\bI_L$.
	$\bXi_X= \textbf{1}_{NL}$, $\bS_X=\textbf{0}_{NL}$,
	$\bXi_H=\textbf{1}_{MN}$, $\bS_H=\textbf{0}_{MN}$.\\
	\textbf{Repeat}
	{
		\begin{algorithmic}[1]
			\State$\overline\bW_X=\hat\bH\herm\hat\bH+ M \bV_H$ \label{alg:UAMPMF_X1}
			\State$[\bC_X,\bD_X]=\text{eig}(\overline\bW_X)$
			\State$\bR_X = \bD_X^{-\frac{1}{2}}\bC_X\herm\hat\bH\herm\bY$,
			$\bPhi_X= \bD_X^{-\frac{1}{2}}\bC_X\herm$
			\State$\bV_{P_X}=|\bPhi_X|^{.2}\bXi_X$
			\State $\bP_X=\bPhi_X\hat\bX-\bV_{\bP_X}\cdot\bS_X$
			\State$\bV_{S_X}=\textbf{1}./(\bV_{P_X}+\hat\lambda^{-1}\mathbf{1})$
			\State$\bS_X=\bV_{S_X}\cdot(\bR_X-\bP_X)$
			\State$\bV_{Q_X}=\textbf{1}./(|\bPhi_X\herm|^{.2}\bV_{S_X})$
			\State$\bQ_X=\hat\bX+\bV_{Q_X}\cdot(\bPhi_X\herm\bS_X)$ \label{alg:UAMPMF_Qx}
			\State{$\bXi_X=\bV_{Q_X}\cdot\bG_X'(\bQ_X,\bV_{Q_X})$}  \label{alg:UAMPMF_VQx}
			\State$\hat\bX=\bG_X(\bQ_X,\bV_{Q_X})$
			\State$\bU_X=\text{diag}(\text{mean}(\bXi_X,2))$
			\State$\overline\bW_H=\hat\bX\hat\bX\herm+ L\bU_X$ \label{alg:UAMPMF:H1} \State$[\bC_H,\bD_H]=\text{eig}(\overline\bW_H)$
			\State$\bR_H = \bD_H^{-\frac{1}{2}}\bC_H\herm\hat\bX\bY\herm$,
			$\bPhi_H = \bD_H^{-\frac{1}{2}}\bC_H\herm$
			\State$\bV_{P_H}=|\bPhi_H|^{.2}\bXi_H\herm$
			\State$\bP_H=\bPhi_H\hat\bH\herm-\bV_{\bP_H}\cdot\bS_H$
			\State$\bV_{S_H}=\mathbf{1}./(\bV_{P_H}+\hat\lambda^{-1}\mathbf{1})$
			\State$\bS_H=\bV_{S_H}\cdot(\bR_H-\bP_H)$
			\State$\bV_{Q_H}=\mathbf{1}./(|\bPhi_H\herm|^{.2}\bV_{S_H})$
			\State$\bQ_H=\hat\bH\herm+\bV_{Q_H}\cdot(\bPhi_H\herm\bS_H)$ \label{alg:UAMPMF:H2}
			\State{$\bXi_H=\bV_{Q_H}\cdot\bG_H'(\bQ_H\herm,\bV_{Q_H}\tra)$}
			\State $\hat\bH=\bG_H(\bQ_H\herm,\bV_{Q_H}\tra)$
			\State$\bU_H=\text{diag}(\text{mean}(\bXi_H,1))$
			\State$\hat\lambda=ML/C$ with $C$ given in \eqref{eq:C}
		\end{algorithmic}
	}
	\textbf{Until terminated}
	\label{alg:UAMPMF}
\end{algorithm}

\section{{System Model and Problem Formulation}}

\subsection{System Model}

Consider a mmWave MIMO system, {where an AP equipped with $M$ antennas is used to serves $U$ user devices, each of which has a single antenna due to the consideration of the size and the cost of user devices. However, we note that the extension of this work to the case of user devices with multiple antennas is straightforward. The number of active devices in a time slot is denoted by $N$, which can be much smaller than $U$. We consider grant-free uplink transmission. 
We assume that the MIMO channel is statistic over $L$ consecutive symbol intervals, which is called a frame. The information bit sequence $\bb_n$ of active user $n$ is mapped to a symbol sequence $\bx_n^T \in \mathbb{C}^{1\times L}$. Then the signal received by the AP over a frame can be represented as
\begin{equation}
  	\bY=\bH\bX+\bW, \label{eqn:Recv1}
\end{equation}
where $\bY=[Y_{ml}]\in \mathbb{C}^{M\times L}$ represents the received signal matrix, $\bX=[\bx_1, ..., \bx_N]^T\in \mathbb{C}^{N\times L}$ denotes the transmitted signal matrix, $\bH=[H_{mn}]\in \mathbb{C}^{M\times N}$ is the MIMO channel matrix 
and $\bW=[W_{ml}]\in \mathbb{C}^{M\times L}$ denotes the additive temporally and spatially white Gaussian noise with zero mean and variance $\sigma_{\omega}^2$.} We assume that no pilot signals are transmitted by the active users, and their IDs are included in the symbol sequences.


\begin{figure}[htbp]
	\centering
	\includegraphics[width=0.5\textwidth]{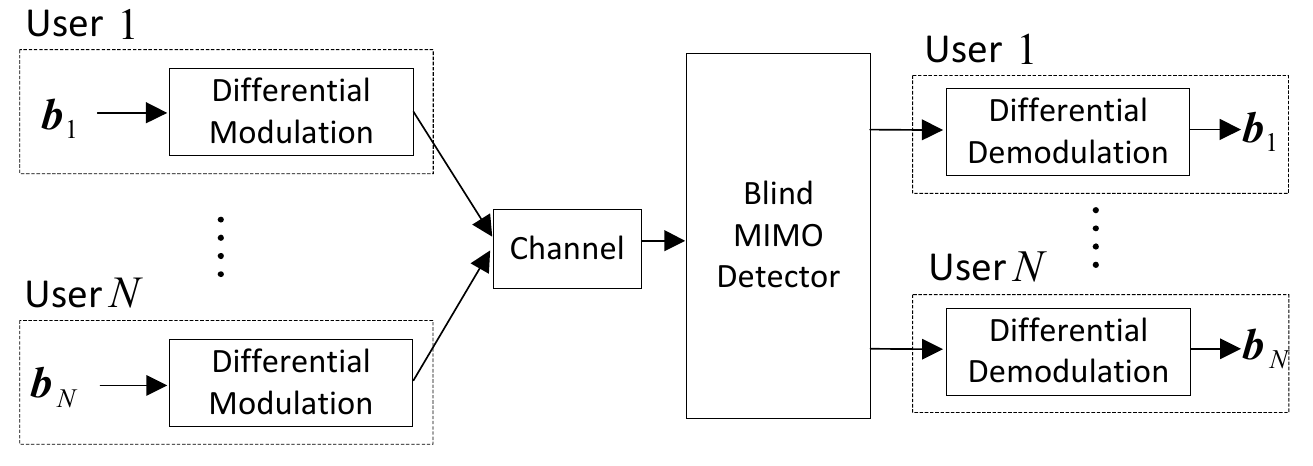}
	\centering
	\caption{Illustration of the system in base-band. }
	\label{fig:encode}
\end{figure}

{The beamspace channel model is widely used due to the highly directional nature of propagation at mmWave frequencies. For simplicity, we assume that the antenna array employed by the AP is a uniform linear one with $M$ antennas. Then the steering vector of the antenna array can be represented as
\begin{equation}
	\ba(\theta)=\left[1, e^{-j2\pi\vartheta},...,e^{-j2\pi\vartheta(M-1)}\right]^T,
\end{equation}
where the normalized spatial angle  $\vartheta=\frac{d}{\lambda'}\sin(\theta)$ with $\theta\in[0,\pi]$ being the physical angle, $d$ denotes the antenna spacing and $\lambda'$ is the wavelength.}
{
The beamspace representation using uniformly spaced spatial angles $\vartheta_k=k/K$ with $k=0:K-1$ can be represented as
\begin{equation}
\bH=\bF\bG \label{eqn:SparseH}
\end{equation}
where $\bF\in \mathbb{C}^{M\times K} (K > M)$ is a partial DFT matrix \cite{tse_viswanath_2005}, i.e., the first $M$ rows of the $K$-point DFT matrix. Due to the limited number of scatterers in a typical mmWave environment, $\bG\in \mathbb{C}^{K\times N}$ is a sparse matrix.
Substituting \eqref{eqn:SparseH} into \eqref{eqn:Recv1} we have
\begin{eqnarray}
\bY&=&\bH\bX+\bW \nonumber \\
   &=&\bF\bG\bX+\bW. \label{eqn:Recv2}
\end{eqnarray}
In this paper, to further reduce the latency of the grant-free random access system and avoid the overhead, we assume that the users do not transmit pilot symbols. 
The task of the receiver at the AP is to directly detect the active users and decode their signals blindly. Specifically,
\begin{itemize}
	\item The matrix $\bF$ is known, but neither $\bG$ nor $\bX$ in \eqref{eqn:Recv2} is known, which need to be estimated jointly based on $\bY$. This is a composite matrix factorization problem. The factorization is possible if we can fully exploit the structures of the matrices, i.e., $\bG$ is a sparse matrix and the entries of $\bX$ are discrete-valued symbols, which are strong constraints for the composite factorization problem.
	\item As the number of active users is unknown, the number of columns of $\bH$ (or the number of rows of $\bX$) \eqref{eqn:Recv2} is unknown, which needs to be determined.
\end{itemize}
In this work, we will design a Bayesian method to recover $\bG$ and $\bX$. Leveraging the UAMP-MF algorithm, an efficient message passing algorithm will be developed.}


\subsection{{Probabilistic Formulation}}

We will develop a Bayesian method to recover $\bG$ and $\bX$ simultaneously, where we specify proper priors for them to reflect the constraints on them. As the entries of matrix $\bG$ are sparse and independent, we use the sparsity-inducing {Bernoulli-Gaussian distribution \cite{Baron2010} as the prior of $\bG$, i.e.,
\begin{equation}
p(\bG) = \prod_{n,k} p(g_{kn})= \prod_{k,n} \left((1-\epsilon)\delta(g_{kn})+ \epsilon \N(g_{kn};0,\nu) \right), \label{eq:priorG}
\end{equation}
where $g_{kn}$ is the $(k,n)$th element of matrix $\bG$, 
$\delta(\cdot)$ denotes the Dirac delta function, and the parameter $\epsilon$ represents the sparsity rate. Here we assume that the number of active users $N$ is known, and its estimation will be discussed later in Section IV.D.
The $n$-th row of $\bX$ stand for the transmitted symbols of the $n$-th active users, which are discrete valued and randomly drawn from the alphabet $\mathcal{A}$. 
So the prior distribution of $\bX$ can be expressed as
\begin{eqnarray}
p(\bX)= \prod_{n,l} p(x_{n,l})=\frac{1}{|\mathcal{A}|}\prod_{n,l} \sum_{a=1}^{|\mathcal{A}|} \delta(x_{n,l}-\alpha_a), \label{priorx}
\end{eqnarray}
where $|\mathcal{A}|$ denotes the size of the alphabet, and $\mathcal{A}=\left\{\alpha_{1},\dots,\alpha_{\left|\mathcal{A}\right|}\right\}$.

As the noise power is normally unknown, so its estimation is also taken into account. We define $\lambda$ as the precision of the noise, i.e., $\lambda=1/\sigma_w^2$.   It is treated as a random variable with an improper prior $f_\lambda(\lambda)\propto 1/\lambda$ \cite{Yuan2021}.
With model \eqref{eqn:Recv2}, we have the following joint conditional distribution and its factorization
\begin{eqnarray}
p(\bX,\bH,\lambda|\bY) \propto p(\bY|\bX,\bH,\lambda)p(\bX)p(\bH|\bG)p(\bG)p(\lambda), 
\label{eqn:factorgraph}
\end{eqnarray}
where
\begin{eqnarray}
p(\bY|\bX,\bH,\lambda) &=& \MN\left(\bY;\bH\bX,\bI_M,\lambda^{-1}\bI_L\right)  \\
p(\bH|\bG)&=& \delta\left(\bH-\bF\bG\right). 
\end{eqnarray}}
{We aim to compute the (proximate) marginals $p(\bx_{n}|\bY)$, }
based on which the transmitted symbols can be detected (at the same time, the channel matrix is also estimated).  However the exact inference is intractable as the involved variables are of high dimension. In this work, we resort to approximate inference techniques, and in particular, with a factor graph representation of the problem, a message passing algorithm will be developed, leveraging the UAMP-MF algorithm.


\section{{ UAMP-MF Based Message Passing Algorithm Design}}


\subsection{{Factor Graph Representation}}

To facilitate the factor graph representation, we define the relevant local functions based on the factorization in \eqref{eqn:factorgraph}:
$f_{\bY}(\bY,\bX,\bH,\lambda) = p(\bY|\bX,\bH,\lambda)$,
$f_{\mathcal{D}}(x_{n,l})=p(x_{n,l})$,
$f_{\bH}(\bH,\bG)=p(\bH|\bG)$,
$f_\lambda(\lambda)=p(\lambda)$, and
$f_{\bG}(\bG)=p(\bG)$.
The local factors and their corresponding probability functions are summarized in Table 1. The factor graph representation of \eqref{eqn:factorgraph} is shown in Fig. \ref{fig:factorgraph},
where we divide the factor graph into two sub-graphs and develop a message passing algorithm in which UAMP-MF is incorporated. Specifically, the UAMP-MF algorithm is mainly used to handle the message passing in Sub-graph I. 

\begin{table}[!hbp]
	\centering
	\caption{Local functions and distributions in \eqref{eqn:factorgraph}}
	\begin{tabular}{lll}
		\hline
		Local Function & Distribution & Function\\
		\hline
		$f_{\bY}(\bY,\bX,\bH,\lambda)$ & $p(\bY|\bX,\bH,\lambda)$ & $\MN\left(\bY;\bH\bX,\bI_M,\lambda\bI_L\right)$\\
		{$f_{\bH}(\bH,\bG)$} & $p(\bH|\bG)$ & $\delta\left(\bH-\bF\bG\right)$ \\
		{$f_{\mathcal{D}}(x_{n,l})$} & $p(x_{n,l})$ & $\frac{1}{|\mathcal{A}|} \sum_{a=1}^{|\mathcal{A}|} \delta(x_{n,l}-\alpha_a)$ \\
        {$f_{\bG}(\bG)$} & $p(\bG)$ & $\prod_{n,k}\left((1-\epsilon)\delta(g_{kn})+\epsilon \N(g_{n,l};0,\nu)\right)$ \\
		$f_\lambda(\lambda)$ & $p(\lambda)$ & $1/\lambda$ \\
		\hline
	\end{tabular}
\end{table}

\begin{figure}[htbp]
	\includegraphics[width=0.5\textwidth]{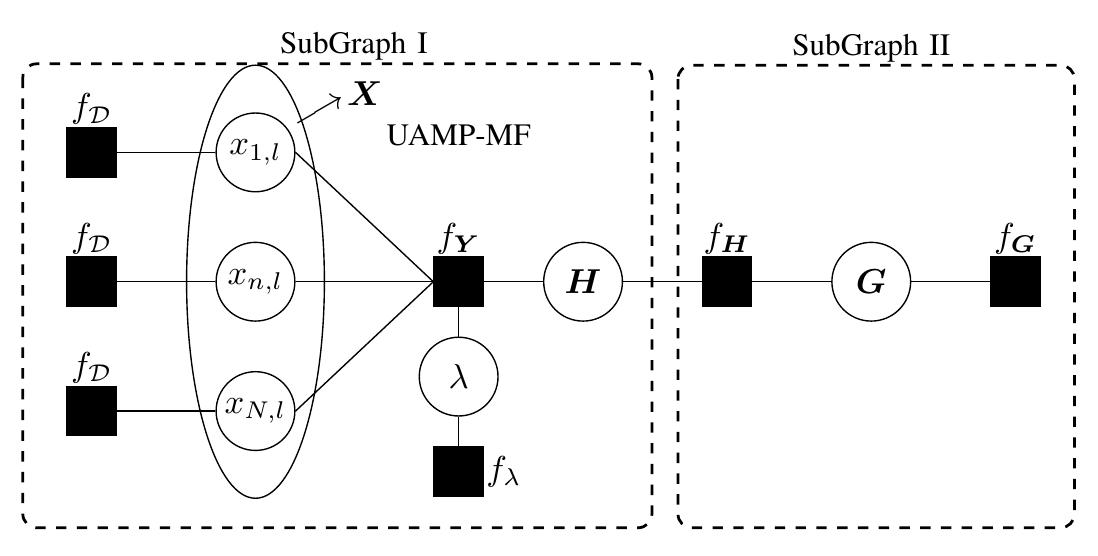}
	\centering
	\caption{Factor graph representation of equation \eqref{eqn:factorgraph}. }
	\label{fig:factorgraph}
\end{figure}

\subsection{{Message Passing Algorithm Design}}

In the derivation of the message passing algorithm, we use the notation $m_{f_{A\to B}}(x)$ to represent the message from (factor or variable) node $A$ to node $B$, which is a function of $x$. It is noted that the message passing algorithm is an iterative one, where each iteration involves a forward recursion and a backward recursion, and the computation of a message in an iteration may require some messages computed in the last iteration. The UAMP-MF based algorithm developed in this work is summarized in Algorithm 3.

\subsubsection{{Message Update in Sub-Graph I}}
According to the derivation of UAMP-MF in \cite{Yuan2022UAMPMF}, we can obtain the backward message about $\bX$ by running Lines 1-9 of Algorithm 2, which produces two matrices $\bQ_X$ and $\bV_{Q_X}$. Due to the decoupling of (U)AMP, we have the message about each entry of $\bX$, i.e.,
\begin{equation}
m_{x_{nl}\to f_{\mD}}= \CN(x_{nl}; q_{x_{nl}},\nu_{q^x_{nl}})
\end{equation}
where $q_{x_{nl}}$ and $\nu_{q^x_{nl}}$ are respectively the $(n,l)$-th element of $\bQ_X$ and $\bV_{Q_X}$. This facilitates the computation of the belief (marginal) of $x_{n,l}$ with the separable prior \eqref{priorx}, i.e., 
\begin{equation}
	b(x_{n,l}) \propto \CN(x_{nl}; q_{x_{nl}},\nu_{q^x_{nl}}) \sum_{a=1}^{|\mathcal{A}|} \delta(x_{n,l}-\alpha_a).
\end{equation}
To make the message passing tractable, we then project $b(x_{n,l})$ to be Gaussian with mean $\hat x_{nl}$ and variance $\nu_{x_{nl}}$. This is equivalent to perform the minimum mean squared error estimation (MMSE), i.e., compute the a posteriori mean and variance of $x_{nl}$, with the following pseudo scalar observation models
\begin{equation}
q_{x_{nl}}=x_{nl}+w_{x_{nl}}, n=1, ...,N, l=1, ..., L \label{eq:scalarmodel}
\end{equation}
where $w_{x_{nl}}$ represents a Gaussian model noise with mean zero and variance $\nu_{q^x_{nl}}$.
It is not hard to show that the \emph{a posteriori} mean $\hat x_{nl}$ and variance $\nu_{x_{nl}}$ of $x_{nl}$ are given by
	\begin{eqnarray}
		\hat x_{nl}&=&\sum\nolimits_{a=1}^{|\mA|}\alpha_a \beta_{nl}^a  \\
		\nu_{x_{nl}}&=&\sum\nolimits_{a=1}^{|\mA|}\beta_{nl}^a|\alpha_a- \hat x_{nl}|^2,
\end{eqnarray}
where
\begin{equation}
	\beta_{nl}^a={\xi_{nl}^a}/{\sum\nolimits_{a=1}^{|\mA|}\xi_{nl}^a}
\end{equation}
with
\begin{equation}
	\xi_{nl}^a=\exp\left({-\nu_{q^x_{nl}}^{-1}|\alpha_a- q_{x_{nl}}|^2}\right).
\end{equation}
The above lead to Lines 1-5 of Algorithm 3.


Then we can stack $\hat x_{nl}$ and $\nu_{x_{nl}}$ to form matrices $\hat\bX=[\hat x_{nl}]$ and $\bXi_X=[\nu_{x_{nl}}]$. We can see that each element $x_{n,l}$ has its own variance. To facilitate subsequent processing, we make an approximation by assuming that the elements of each row in $\hat\bX$ share a same variance, which is the average of their variances. Then they can be collectively characterized by a matrix normal distribution, i.e., $b(\bX)=\MN(\bX; \hat\bX, \bU_X, \bV_X)$ with $\bU_X=\text{diag}(\text{mean}(\bXi_X,2))$ and $\bV_X=\bI_L$, where  $\text{mean}(\bXi_X,2)$ represents the average operation on the rows of $\bXi_X$. These lead to Lines 6 and 7 of Algorithm 3.

\subsection{{Message Update in Sub-Graph II}}

\begin{algorithm}
	\caption{UAMP-MF Based Blind Joint UACESD}
	\textbf{Initialization}: $\hat N=U_r$, $\bU_H=\bI_M$,  $\hat\bH=\mathcal{R}_c(M,\hat N)$, $\bV_X=\bI_L$,
	$[\bU_F, \bLambda_F, \bV_F]=\text{SVD}(\bF) $, $\bPhi_G=\bU^H_F\bF$, and $\bLambda_G=  (\bLambda_F\bLambda_F\herm) \boldsymbol{1}_{K}\boldsymbol{1}_N\tra$. \\
	\textbf{Repeat}
	\begin{algorithmic}[1]
		\State Compute $\bV_{Q_X}$ and $\bQ_X$ with Lines 1-9 in Algorithm \ref{alg:UAMPMF}
		\State 	{$\forall n,l,a: \xi_{nl}^a=\exp\left({-\nu_{q^x_{nl}}^{-1}|\alpha_a- q_{x_{nl}}|^2}\right)$} \label{alg:xi}
		\State 	$\forall n,l,a: \beta_{nl}^a={\xi_{n,l}^a}/{\sum\nolimits_{a=1}^{|\mA|}\xi_{nl}^a}$
		\State 	$\forall n,l: \hat  x_{nl}=\sum\nolimits_{a=1}^{|\mA|}\alpha_a \beta_{nl}^a$
		\State 	$\forall n,l: \nu_{x_{nl}}=\sum\nolimits_{a=1}^{|\mA|}\beta_{nl}^a|\alpha_a- \hat x_{nl}|^2$ \label{alg:vx}
		\State  $\hat \bX=\big[\hat  x_{nl}\big]$, $\bXi_X=\big[\nu_{x_{nl}}\big]$
		\State$\bU_X=\text{diag}(\text{mean}(\bXi_X,2))$ \label{alg:UAMPX_end}
		{\State Compute $\bV_{Q_H}$ and $\bQ_H$ with Lines \ref{alg:UAMPMF:H1} - \ref{alg:UAMPMF:H2} in Algorithm \ref{alg:UAMPMF}
		\State $\bR_G=\bU^H_F \bQ_{\bH}\herm$,
        \State {$\tau=\left< 1./\bV_{Q_H}\tra\right>$} \label{alg:UAMPG_start}
		\State $\bV_{P_G} = \bLambda_G \cdot \bV_{G}$
		\State $\bP_G= \bPhi_G\hat\bG - \bV_{P_G} \cdot  \bS_G$
		\State $\bV_{S_G}= 1./ (\bV_{P_G}+\tau^{-1}) $
		\State $\bS_G= \bV_{S_G} \cdot (\bR_G-\bP_G) $
		\State $1./\bV_{Q_G}=\boldsymbol{1}_K \left(\frac{1}{K}\boldsymbol{1}_M^T(\bLambda_G\cdot \bV_{S_G})\right)$
		\State $ \bQ_G = \hat\bG + \bV_{Q_G}  \cdot (\mathbf{\Phi}_G^H \bS_G)$ \label{alg:UAMPG_end}
		\State $\forall n,k$ update $\hat g_{kn}$ and $\nu_{g_{kn}}$ by \eqref{eq:b_g}
		\State stack $\hat g_{kn}$ and $\nu_{g_{kn}}$ into matrices $\hat\bG$ and $\bV_G$
		\State update sparsity rate $\epsilon$ by \eqref{eq:epsilon}}
		\State $\bXi_H=\left|\bU_F\right|^2\left(1./(\tau+1./\bV_{P_G})\right)$
		\State  $\hat\bH= \bU_F\left(\bV_H \cdot (\bR_G\cdot \tau+ \hat\bP_G./\bV_{P_G})\right)$	\label{alg:UAMP_HatH}        \State$\bV_H=\text{diag}(\text{mean}(\bXi_H,1))$  \label{alg:UAMPH_end}
		\State$\hat\lambda=ML/C$ with $C$ given in \eqref{eq:C}
	\end{algorithmic}
		\textbf{Until terminated}
	\label{alg:UACESD}
\end{algorithm}

{According to the derivations of the UAMP-MF algorithm, the outgoing message of Sub-graph I is about matrix $\bH$, which is also decoupled Gaussian distributions with mean matrix $\bQ_{\bH}\in \mathbb{C}^{M\times N}$ and variance matrix $\bV_{Q_H}\in \mathbb{C}^{M\times N}$ (the entries in the matrix are the variances of the corresponding entries in $\bH$), which can be computed using Lines 13-21 of Algorithm \ref{alg:UAMPMF}.

Next, we need to compute the message passed from the factor node $f_{H}$ to variable node $\bG$. Considering the hard constraint $f_{H}(\bH,\bG)=\delta(\bH-\bF\bG)$, matrix $\bQ_{\bH}$ can be regarded as an observation matrix with the following pseudo model
\begin{eqnarray}
\bQ_{\bH}=\bF\bG+\bW_H \label{eq:pseudoH3}
\end{eqnarray}
where $\bW_H$ is a model noise matrix, whose entries are independent with means zero and variances given by the corresponding entries in $\bV_{Q_H}$. To enable low complexity inference, we make an approximation that the model noise is white, and the entries have a same precision $\tau$ (or variance $\tau^{-1}$), which is the average of the individual precisions in the matrix $1./\bV_{Q_H}$, i.e., $\tau=\left< 1./\bV_{Q_H}\tra\right>$. After doing this, \eqref{eq:pseudoH3} becomes a standard linear problem with known system transfer matrix $\bF$, hence UAMP in Algorithm 1 can be incorporated to solve it.

We choose UAMPv2 due to its lower complexity. It is noted that UAMPv2 in Algorithm 1 is developed for a single vector problem. Next, we extend it to the case for a matrix problem in \eqref{eq:pseudoH3}. With the SVD
$\bF=\bU_F \bLambda_F \bV_F$, 
we carry out a unitary transformation with $\bU^H_F$ to \eqref{eq:pseudoH3}, yielding
\begin{equation}
	\bR_G=\bPhi_G \bG + \bW_{H}',
\end{equation}
where
\begin{eqnarray}
\bR_G=\bU^H_F \bQ_{\bH}\in \mathbb{C}^{M\times N}\label{eq:bR_G}, \\
\bPhi_G=\bU^H_F\bF \in \mathbb{C}^{M\times K} \label{eq:bPhi_G},
\end{eqnarray}
and $\bW_{H}'=\bU^H_F \bW_{H}$. Since $\bF$ is a unitary matrix, $\bW_{H}'$ is still a white Gaussian noise matrix and has the same variance as $\bW_{H}$.
Matrix $\bLambda_F$ is an $M\times K$ rectangular diagonal matrix. In Algorithm 1, a vector $\boldsymbol{\lambda}=\bLambda\bLambda^H\boldsymbol{1}$ is defined. To make the subsequent expression concise, we define a matrix $\bLambda_G\in \mathbb{C}^{M\times N}$ as
\begin{eqnarray}
\bLambda_G=  \boldsymbol{\lambda}_F\boldsymbol{1}_N\tra \label{eq:bPsi_G}
\end{eqnarray}
where $\boldsymbol{\lambda}_F=(\bLambda_F\bLambda_F\herm) \boldsymbol{1}_{K}$. 

Following UAMPv2 in Algorithm 1, we can compute the mean and variances about the entries in matrix $\bG$ in the following.
Firstly, we compute matrices $\bV_{P_G}$ and $\bP_G$ as
\begin{eqnarray}
\bV_{P_G} &=& \bLambda_G \cdot \bV_{G} \\
\bP_G &=& \bPhi_G\hat\bG - \bV_{P_G} \cdot \bS_G \label{eq:PG}
\end{eqnarray}
where $\hat\bG$ and $\bV_{G}$ represent the mean matrix and variance matrix. The matrices are obtained based on the a posteriori distribution $b(\bG)$, which are computed in \eqref{eq:hatG}. It is noted that, before the first iteration, we use the initialization $\hat\bG=\boldsymbol{0}_{K\times N}$ and $\bV_{G}=\boldsymbol{1}_{K\times N}$. In \eqref{eq:PG} matrix $\bS_G$ is initiated as 0 for the first iteration, and is updated in $\eqref{eq:hatSG}$.

{Then, we update intermediate matrices $\bV_{S_G}$ and $\bS_G$ as
\begin{eqnarray}
\bV_{S_G}=1./ (\bV_{P_G}+\tau^{-1}{\boldsymbol{1}_{M\times N}})\\
\bS_G= \bV_{S_G} \cdot (\bR_G-\bP_G), \label{eq:hatSG}
\end{eqnarray}
where $\tau$ represents the model noise precision. 
According to Line 5 of Algorithm 1, we can get $\bV_{Q_G}$ and $\bQ_G$ as
\begin{eqnarray}
1./\bV_{Q_G}&=&\boldsymbol{1}_K \left(\frac{1}{K}\boldsymbol{1}_M^T(\bLambda_G\cdot \bV_{S_G})\right)\\
\bQ_G &=& \hat\bG + \bV_{Q_G}  \cdot (\mathbf{\Phi}_G^H \bS_G).
\end{eqnarray}
The above lead to Lines 9-16 of Algorithm 3.

Due to the decoupling of (U)AMP, we have the following scalar pseudo observation models
\begin{equation}
g_{kn}=q_{g_{kn}}+w_{kn} \label{eq:pesudo_g}
\end{equation}
where $q_{g_{kn}}=\left[\bQ_G\right]_{kn}$, $\nu^q_{g_{kn}}=\left[\bV_{Q_G}\right]_{kn}$, and $w_{kn}$ represents the Gaussian noise with variance $\nu^q_{g_{kn}}$.}
Then, we can compute the a posteriori distributions of $\{g_{kn}\}$  with the pseudo models and the priors $\{p(g_{kn})\}$.
As discussed previously,  matrix $\bG\in \mathbb{C}^{K\times N}$ is sparse and we employ the sparsity promoting Bernoulli-Gaussian prior in \eqref{eq:priorG}. The a posteriori distribution $b(g_{kn})$ can be obtained as
\begin{eqnarray}
&&b(g_{kn})=\frac{p(g_{kn})\CN(g_{kn}; q_{g_{kn}},\nu^q_{g_{kn}})}{\int_{g_{kn}} p(g_{kn})\CN(g_{kn}; q_{g_{kn}},\nu^q_{g_{kn}})}\nonumber\\
&&=\frac{(1-\epsilon)\CN(q_{g_{kn}};0,\nu^q_{g_{kn}})\delta(g_{kn})+\alpha_{kn}\CN(g_{kn};\gamma_{kn},\nu_{\gamma_{kn}})}
{(1-\epsilon)\CN(\hat q_{g_{kn}};0,\nu^q_{g_{kn}})+ \alpha_{kn}}\nonumber\\
&&=(1-\pi_{kn})\delta(g_{kn})+\pi_{kn}\CN(g_{kn};\gamma_{kn},\nu_{\gamma_{kn}})
\end{eqnarray}
where
\begin{eqnarray}
		&&\gamma_{kn}= \frac{ q_{g_{kn}} \nu}{\nu^q_{g_{kn}}+\nu}, 
		\nu_{\gamma_{kn}}=\frac{\nu^q_{g_{kn}}\nu}{\nu^q_{g_{kn}}+\nu} \nonumber\\
		&&\alpha_{kn}=\epsilon \CN( q_{g_{kn}};0,\nu^q_{g_{kn}}+\nu)\nonumber\\
		&&\pi_{kn}=\frac{\alpha_{kn}}{(1-\epsilon)\CN( q_{g_{kn}};0,\nu^q_{g_{kn}})+\alpha_{kn}}\label{eq:pi}.
\end{eqnarray}
To avoid intractable message passing due to the complex form of the belief $b(g_{kn})$, we project $b(g_{kn})$ to a Gaussian form, i.e.,
\begin{eqnarray}
\text{Proj}\left(b(g_{kn})\right)=\CN(g_{kn}; \hat g_{kn}, \nu_{g_{kn}}) \label{eq:b_g}
\end{eqnarray}
where
\begin{eqnarray}
\hat g_{kn} &=& \pi_{nk} \gamma_{kn}, \\
\nu_{g_{kn}} &=& \pi_{nk}^2\nu_{\gamma_{kn}}.
\end{eqnarray}
The above lead to Line 17 of Algorithm 3.

We stack $\hat g_{kn}, \nu_{g_{kn}}\forall n,k$ into matrices $\hat \bG$ and $\bV'_G$,
\begin{eqnarray}
\hat \bG=\{\hat g_{kn}\},
\bV'_G=\{\nu_{g_{kn}}\}.\label{eq:hatG}
\end{eqnarray}
According to Line 7 of Algorithm 1, we need to average the variance matrix $\bV'_G$ column wise, i.e.,
\begin{equation}
\bnu_g^T=\frac{1}{K}\boldsymbol{1}_K\tra\bV'_G
\end{equation}
then we get variance matrix $\bV_G$ by
\begin{equation}
\bV_G=\boldsymbol{1}_K \bnu_g^T.
\end{equation}

As the knowledge about the sparsity rate $\epsilon$ is often unknown, we can also incorporate its estimation into the message passing algorithm,  with the expectation maximization framework as in \cite{Vila2013}, i.e.,
\begin{equation}
\epsilon=\mathop{\text{arg max}}\limits_{\epsilon\in(0,1)}\sum_{k,n}\int_{g_{kn}}\Big(\ln p(g_{kn}|\epsilon)b(g_{kn})\Big),
\end{equation}
and the value of $\epsilon$ can be obtained by letting the derivative be zero, i.e.,
\begin{eqnarray}
&&\frac{\partial}{\partial\epsilon}\sum_{k,n}\int_{g_{kn}}\Big(\ln p(g_{kn}|\epsilon)b(g_{kn})\Big) \nonumber\\
&&\ \ \ \ =\sum_{k,n}\int_{g_{kn}}\Big(\frac{\partial}{\partial\epsilon}\ln p(g_{kn}|\epsilon)b(g_{kn})\Big)=0 \label{eq:EM}
\end{eqnarray}
With the Bernoulli-Gaussian distribution,
\begin{eqnarray}
\frac{\partial}{\partial\epsilon}\ln p(g_{kn}|\epsilon)
&=&\frac{-\delta(g_{kn})+ \N(g_{kn};0,\nu)}{(1-\epsilon)\delta(g_{kn})+ \epsilon \N(g_{kn};0,\nu)}\nonumber\\
&=&
\begin{cases}
1/\epsilon,& g_{kn}\neq 0 \\
-1/(1-\epsilon),& g_{kn}= 0 \label{eq:dev_ln_g}
\end{cases}
\end{eqnarray}
Substituting \eqref{eq:b_g} and \eqref{eq:dev_ln_g} into \eqref{eq:EM}, we can obtain that \cite{Vila2013}}

\begin{eqnarray}
\epsilon=\frac{1}{KN}\sum\nolimits_{k,n}\pi_{kn}. \label{eq:epsilon}
\end{eqnarray}
The above correspond to Lines 17-19 of Algorithm 3.

Finally, we can compute the a posteriori distribution $b(\bH)$, whose entries are approximated to be independent and Gaussian with variances and means given by the corresponding entries of the matrices  
\begin{eqnarray}
\bXi_H&=&\left|\bU_F\right|^2\left(1./(\tau+1./\bV_{P_G})\right), \\
\hat\bH&=&\bU_F\left(\bV_H \cdot (\bR_G\cdot \tau+ \hat\bP_G./\bV_{P_G})\right).
\end{eqnarray}
Similar to the case for updating $b(\bX)$, 
to accommodate $\{b(h_{mn})\}$ with a matrix normal distribution, we make an approximation by assuming the entries in each column of $\bH$ share a common variance, which is the average of their variances. Then $b(\bH)=\MN(\bH;\hat\bH,\bU_H,\bV_H)$ with  $\bV_H=\text{diag}(\text{mean}(\bXi_H,1))$ and $\bU_H=\bI_M$, where $\text{mean}(\bXi_H,1)$ represents the average operation on the columns of $\bXi_H$. After that the noise precision is updated. The above lead to Lines 20-23 of Algorithm 3.


\subsection{{Estimation of the Number of Active Users}}
{In the above, we assume the number of active users $N$ is known, which needs to be estimated in a practical scenario.   In the blind UACESD algorithm, the number of active users, $N$, corresponds to the number of rows in $\bX$ or the number of columns in $\bH$. We follow the method in {\cite{Parker2014}} to estimate $N$ in an iterative manner based on the singular values of the estimated matrix $\hat\bH$. Initially, we set it to the maximum number of possible active users $\bar N$, and update its value in an iterative manner.
With the sorted singular values $\sigma_n$ of the estimated matrix $\hat\bH$, we compute the pairwise ratios $R_n=\sigma_n/\sigma_{n+1}$. The rank $N$ is estimated based on the largest ratio, i.e.,
\begin{eqnarray}
\hat N =\text{arg max}_n R_n
\end{eqnarray}
which is the largest gap in successive singular values. However, this estimate candidate is accepted only if this maximizing ratio exceeds the average ratio,
\begin{eqnarray}
\bar R= \frac{1}{\hat N-2}\sum\nolimits_{i\neq \hat N} R_i.
\end{eqnarray}
If the candidate $\hat N$ is accepted, the matrices $\bH$ and $\bX$ are pruned and then the blind UACESD algorithm is run. If the candidate is not accepted, the blind UACESD algorithm is run for one more iteration, and the estimation of  $N$ is performed until a candidate $\hat N$ is accepted.}  

\subsection {{Handling Ambiguity}}
{An inherent problem with the matrix factorization problem in \eqref{eqn:Recv2} is the ambiguity. Due to the strong constraints on matrices $\bH$ and $\bX$, there is only a phase ambiguity. To solve the problem, we have two approaches. One approach is to use differential modulations for all users, so the phase ambiguity does not have impact on demodulation. The other approach is that all users transmit few pilot symbols, which are known to the AP, so that the ambiguity can be mitigated. As the number of pilot symbols can be very small, the overhead can be ignored. In this paper, we use the first approach in the simulations.}

\section{Simulation Results}

{In this section, we evaluate the performance of the UAMP-MF based blind joint UASD algorithm.
We assume that the access point is equipped with $M=100$ antennas and serves $U=300$ users in an area. In most of the simulations, we assume that the rate of active users is $10\%$ and $16.7\%$, i.e., 30 and 50 active users during a transmission block. The size of matrix $F$ is $100 \times 150$, the number of paths per user is 10, and the signals arrive at the AP with random angles. {Differential QPSK} modulation is employed. Both coded system and uncoded system are considered. For the coded system, we use a rate $1/2$ convolutional code with generator polynomial $[5,7]_8$.  The simulation results are obtained by averaging over $10^{5}$ Monte Carlo trials.
The signal to noise power ration (SNR) is defined as
\begin{equation}
	\text{SNR}=\frac{||\bH\bX||^2 / MNL}{\sigma^2},
\end{equation}
where the numerator is the power of the signal per antenna per user, and $\sigma^2$ is the power of noise. To examine the performance of user activity detection, we define active user error rate (AER) as
\begin{eqnarray}
	\text{AER}=\frac{\text{\#of active users}-\text{\#of active users identified successfully}}
	{\text{\#of active users}}. \nonumber
\end{eqnarray}}

{To the best of our knowledge, there are no existing algorithms in the literature that can be used to solve the blind joint UACESD problem formulated in this paper. Hence, in this section, we compare the joint blind UACESD algorithm (named 'Blind-UACESD') with some performance bounds.
The first bound is the performance of the system where the number of active users $N$ is known, and joint channel estimation and signal detection is performed, which is denoted by 'Blind-CESD'. The second one is the performance of the system where the active user number $N$ is known and a frame of transmitted signal $\bX$ is used as pilot to estimate the channel, which is named as 'CESD'. In the last one, we assume the perfect channel state information, i.e., the channel matrix $\bH$ is known, so the bilinear problem is degenerated to a linear problem.
The system in this case is denoted by 'SD'. The bit error rate (BER), frame error rate (FRE), the normalized mean squared error (NMSE) of channel estimation, and AER are used to evaluate the performance. }
	

We first consider an uncoded system.  Figures \ref{fig:Uncode_BER} and \ref{fig:Uncode_FER}  show the BER and FER performance of the proposed algorithm and the relevant performance bounds. We set the number of active users to $N=30$ and 50, and the length of the frame $L=200$. From Fig. \ref{fig:Uncode_BER} we can see that the BER performance of Blind-UACESD and Blind-CESD are almost the same, and Blind-UACESD performs very well and there is about 1dB performance gap compared with CESD and SD. We can also see that, when the number of active users is increased to 50, Blind-UACESD exhibits certain performance loss in the relatively lower SNR range. This is because it is difficult to identify the number of active users in the low SNR range.  Figure \ref{fig:Uncode_FER} shows the FER  performance of the system versus SNR, from which the same observations as Fig. \ref{fig:Uncode_BER} are made.  Figure \ref{fig:Uncode_FER} also shows the AER performance of  Blind-UACESD when the number of active users is 50. We see that the user activity detection performs well, and AER was not observed in the simulations when SNR is above 3dB.

\begin{figure}[!t]
	\centering
	\includegraphics[width=0.4\textwidth]{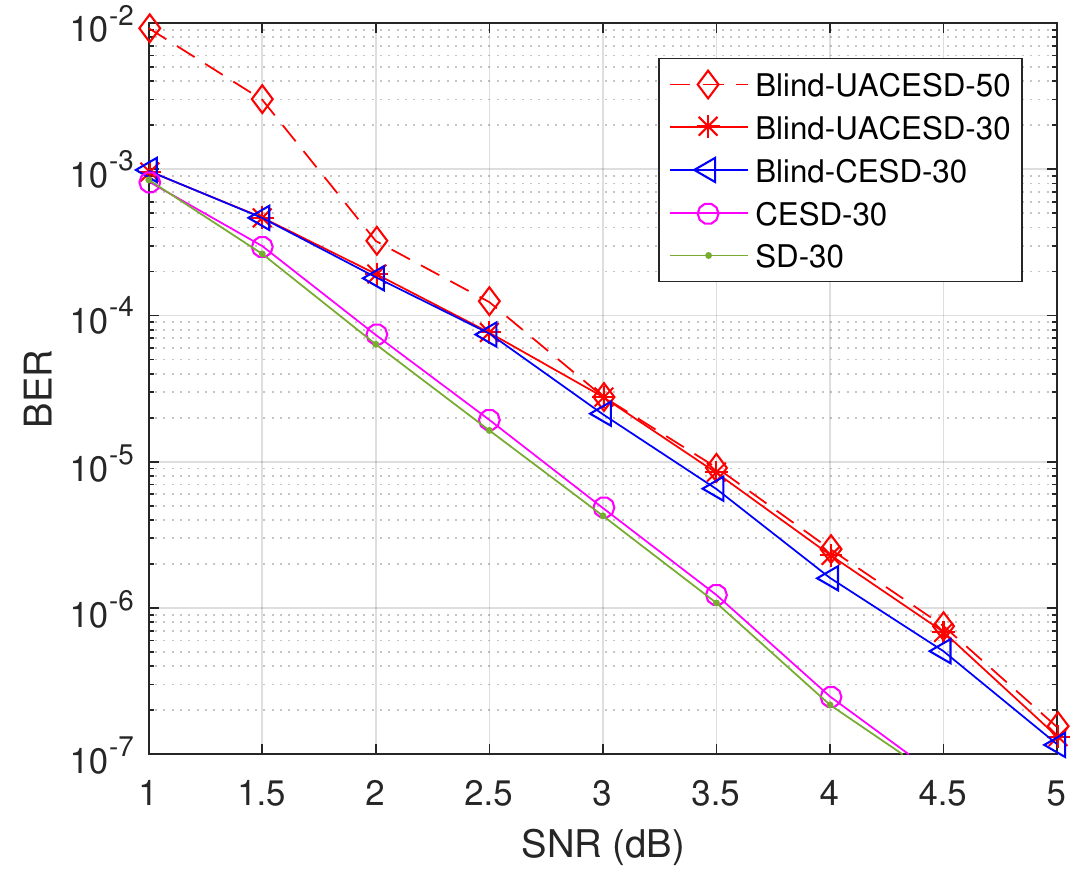}
	\centering
	\caption{{BER performance of the system, where $L=200$ and $N=30$ and $50$.}}
	\label{fig:Uncode_BER}
\end{figure}

\begin{figure}[!t]
	\centering
	\includegraphics[width=0.4\textwidth]{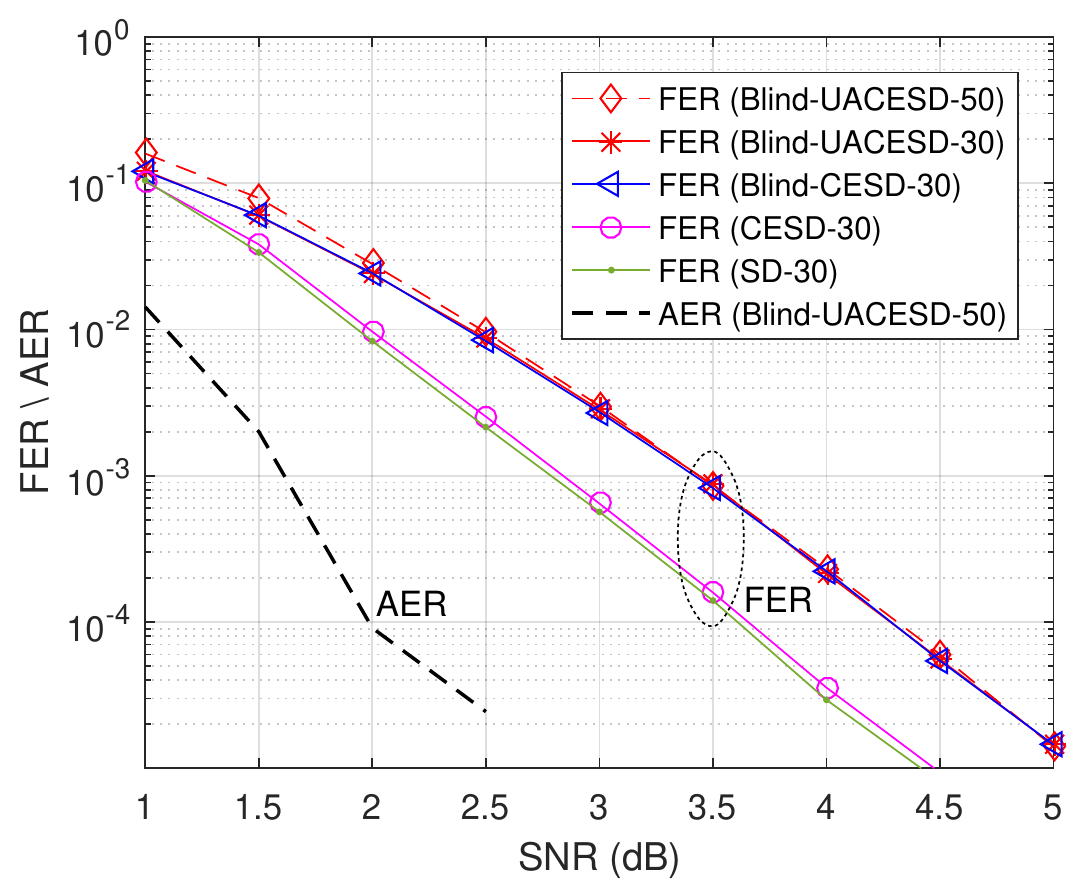}
	\centering
	\caption{{FER performance in an uncoded system, where $L=200$ and $N=30$ and 50.}}
	\label{fig:Uncode_FER}
\end{figure}


In Fig.\ref{fig:VsN}, we vary the number of active users from 10 to 60 and examine the performance of the system when SNR =3dB and 4dB. It can be observed that, the BER performance of the system increases slightly with the number of active users. When SNR=3dB and the number of active users is 60, the algorithm experiences difficulties in active user detection, thereby leading to higher BER. In general, the system with the proposed Blind-UACESD performs well, and its performance is close to those performance bounds.
To examine the channel estimation performance, the performance of NMSE versus SNR is shown in Fig. \ref{fig:Uncode_MSE}, where the ambiguity is removed. We can be seen from the figure that, the NMSE of channel estimation of Blind-UACESD is almost the same as that obtained using pilot signals (i.e., CESD). 
Again, when the number of active users is 50, there is also performance loss in the low SNR range, compared with the case when the number of active users is 30.

\begin{figure}[!t]
	\centering
	\includegraphics[width=0.4\textwidth]{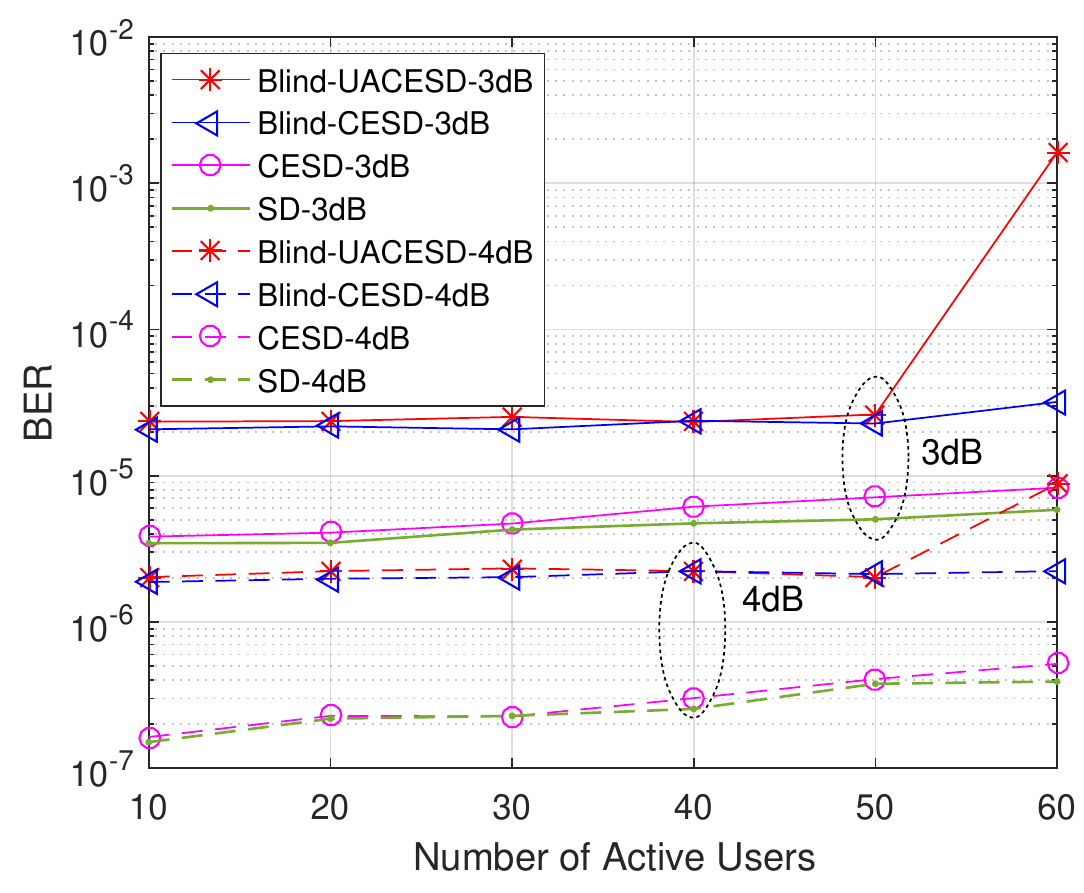}
	\centering
	\caption{{BER performance versus the number of active users, where SNR = 3dB and 4dB.}}
	\label{fig:VsN}
\end{figure}

\begin{figure}[!t]
	\centering
	\includegraphics[width=0.4\textwidth]{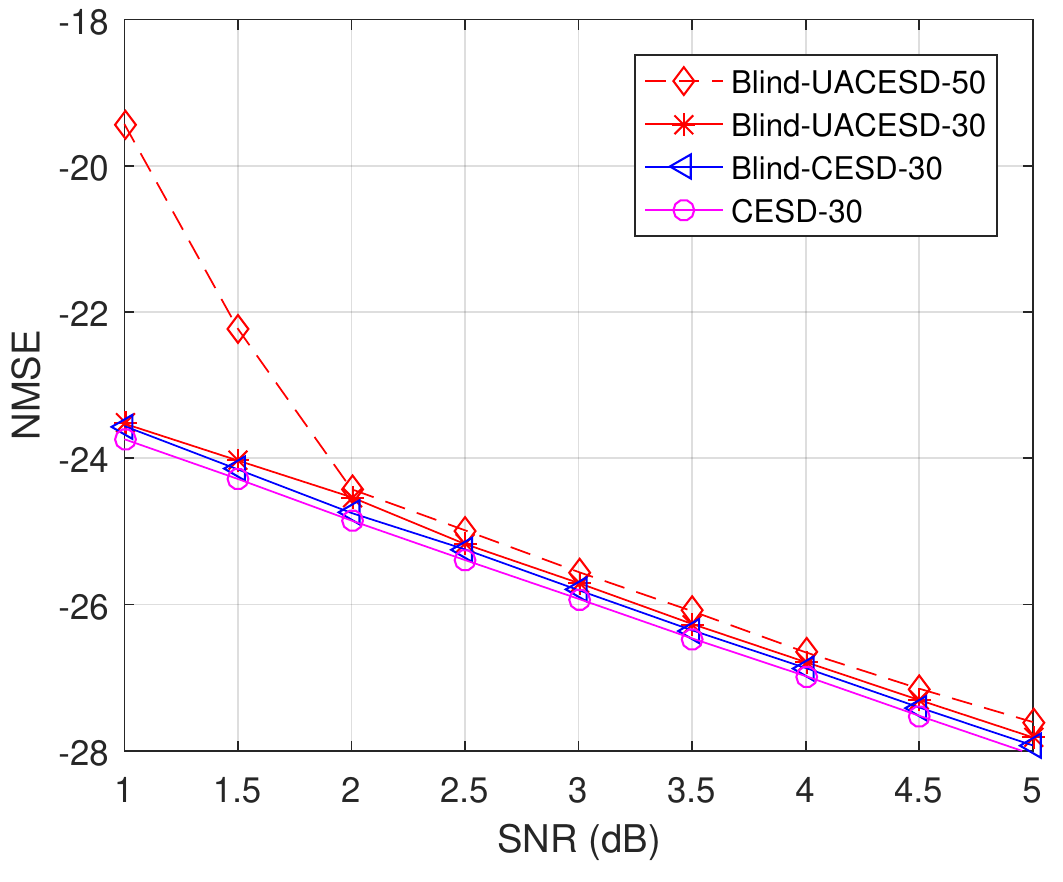}
	\centering
	\caption{{MSE performance of channel estimation in an uncoded system, where $L=200$ and $N=30$ and 50. }}
	\label{fig:Uncode_MSE}
\end{figure}

Then we consider a coded system where all the users employ a rate-$1/2$ convolutional code with generator polynomial $[5,7]_8$. Figures \ref{fig:Code_BER} and \ref{fig:Code_FER} show the BER, FER, and AER performance of the system, respectively. The frame length $L=300$. We can see that, compared to the results in the uncoded case, the performance of the coded system is improved considerably. 


In the design of the Blind-UACESD algorithm, the variance (reciprocal of the precision) of the noise is assumed unknown, and its estimation is incorporated in the algorithm. Lastly, we examine the performance of noise variance estimation and the results are shown in Fig. \ref{fig:Code_Noise} where we vary the SNR. We can see that the estimated noise variance matches the real noise variance very well at different SNRs.

\begin{figure}[!t]
	\centering
	\includegraphics[width=0.4\textwidth]{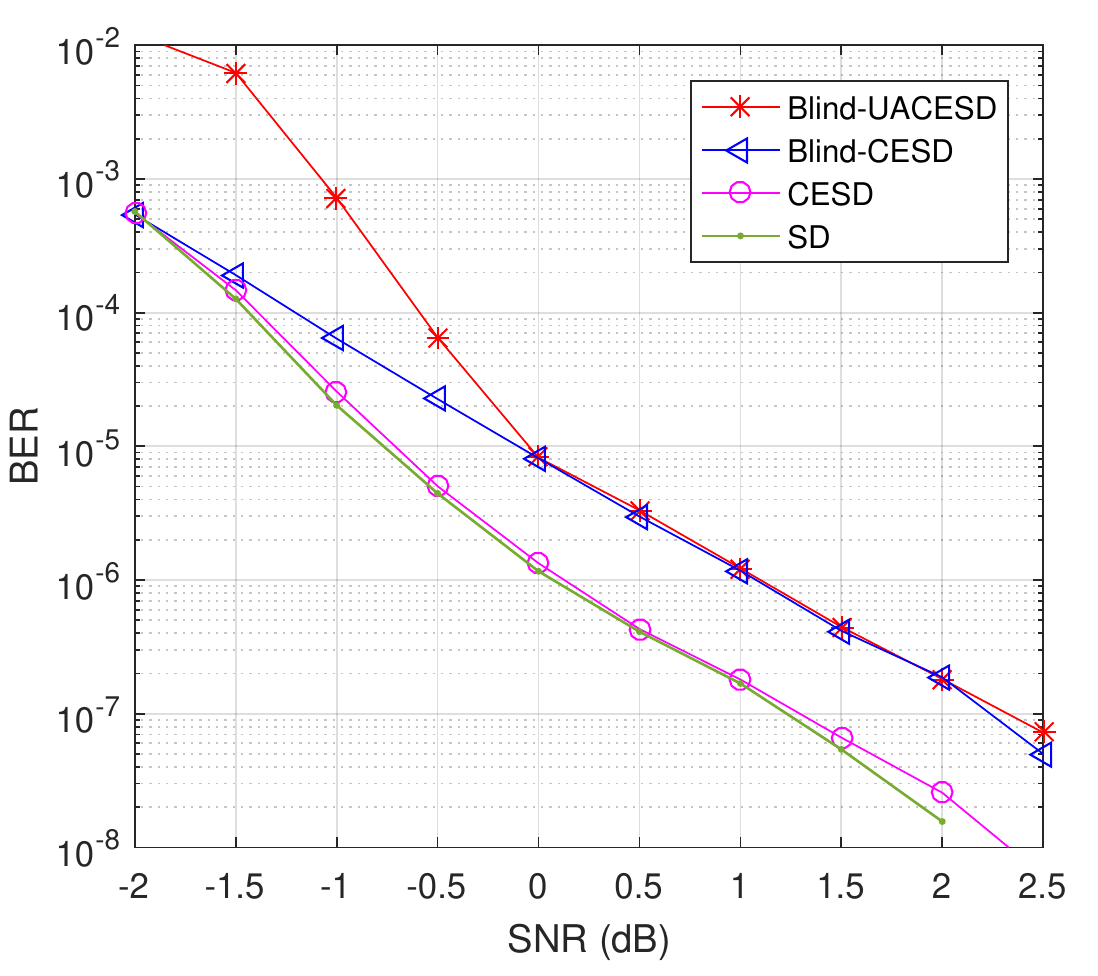}
	\centering
	\caption{{BER performance with a coded system, where $L=300$ and {$N=30$ and 50}.}}
	\label{fig:Code_BER}
\end{figure}
\begin{figure}[!t]
	\centering
	\includegraphics[width=0.4\textwidth]{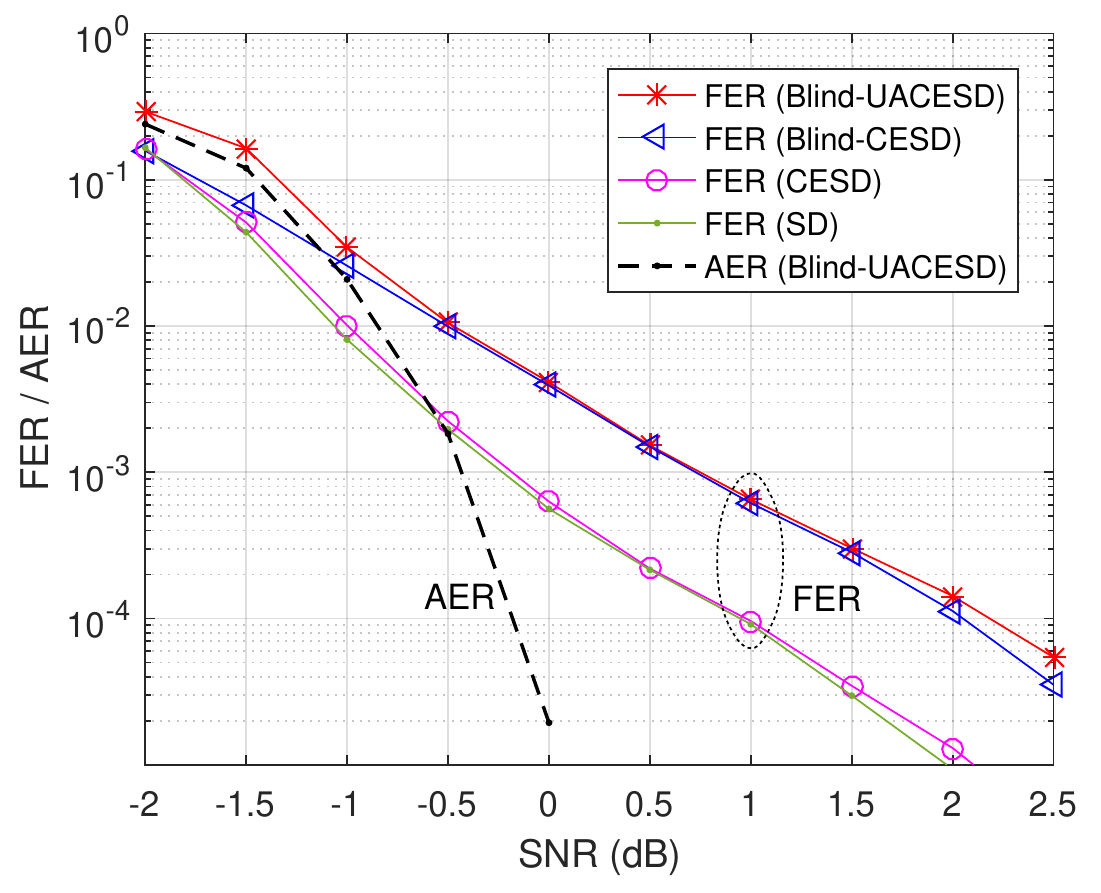}
	\centering
	\caption{{FER performance with a coded system, where $L=300$ and $N=30$.}}
	\label{fig:Code_FER}
\end{figure}

\begin{figure}[!t]
	\centering
	\includegraphics[width=0.4\textwidth]{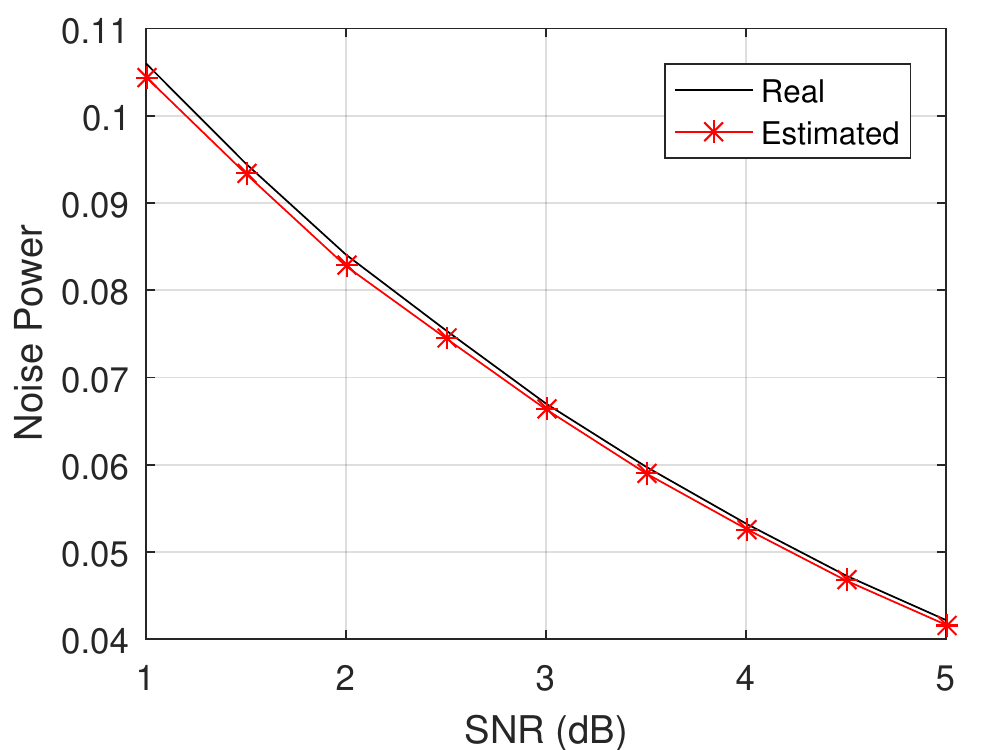}
	\centering
	\caption{Noise variance estimation versus SNR.}
	\label{fig:Code_Noise}
\end{figure}

\section{Conclusions}
{In this paper, we have investigated the issue of joint user activity detection, channel estimation and signal detection without the use of pilot signals in mmWave MIMO-based grant-free random access to further reduce the overhead and latency. The blind joint user activity detection, channel estimation and signal detection is formulated as a contained composite matrix factorization problem, and we have shown that it can be solved by exploiting the structures of the factoring matrices, i.e., the channel matrix is sparse in the beam space domain and the entries of the data matrix are discrete valued. Leveraging UAMP-MF, we have developed a message passing Bayesian algorithm to solve the problem. Extensive simulation results are provided to demonstrate the effectiveness of the grant-free random access scheme without the use of pilot signals.}


\bibliographystyle{IEEEtran}
\bibliography{bibliography}

\begin{thebibliography}{10}
\providecommand{\url}[1]{#1}
\csname url@samestyle\endcsname
\providecommand{\newblock}{\relax}
\providecommand{\bibinfo}[2]{#2}
\providecommand{\BIBentrySTDinterwordspacing}{\spaceskip=0pt\relax}
\providecommand{\BIBentryALTinterwordstretchfactor}{4}
\providecommand{\BIBentryALTinterwordspacing}{\spaceskip=\fontdimen2\font plus
\BIBentryALTinterwordstretchfactor\fontdimen3\font minus
  \fontdimen4\font\relax}
\providecommand{\BIBforeignlanguage}[2]{{%
\expandafter\ifx\csname l@#1\endcsname\relax
\typeout{** WARNING: IEEEtran.bst: No hyphenation pattern has been}%
\typeout{** loaded for the language `#1'. Using the pattern for}%
\typeout{** the default language instead.}%
\else
\language=\csname l@#1\endcsname
\fi
#2}}
\providecommand{\BIBdecl}{\relax}
\BIBdecl

\bibitem{Lee1999Nature}
D.~D. Lee, ``Learning parts of objects by non-negative matrix factorization,''
  \emph{Letter of Nature}, vol. 401, pp. 788--791, 1999.

\bibitem{Rubinstein2010DL}
R.~{Rubinstein}, A.~M. {Bruckstein}, and M.~{Elad}, ``Dictionaries for sparse
  representation modeling,'' \emph{Proceedings of the IEEE}, vol.~98, no.~6,
  pp. 1045--1057, June 2010.

\bibitem{Zhu2011CSMU}
H.~{Zhu}, G.~{Leus}, and G.~B. {Giannakis}, ``Sparsity-cognizant total
  least-squares for perturbed compressive sampling,'' \emph{IEEE Transactions
  on Signal Processing}, vol.~59, no.~5, pp. 2002--2016, May 2011.

\bibitem{Cand2011}
\BIBentryALTinterwordspacing
E.~J. Cand{e}s, X.~Li, Y.~Ma, and J.~Wright, ``Robust principal component
  analysis?'' \emph{J. ACM}, vol.~58, no.~3, Jun. 2011. [Online]. Available:
  \url{https://doi.org/10.1145/1970392.1970395}
\BIBentrySTDinterwordspacing

\bibitem{yuan2021BiUTAMP}
Z.~Yuan, Q.~Guo, and M.~Luo, ``Approximate message passing with unitary
  transformation for robust bilinear recovery,'' \emph{IEEE Trans. Signal
  Process.}, vol.~69, pp. 617--630, Dec. 2020.

\bibitem{Lee2001MIT}
D.~D. Lee, ``Algorithms for non-negative matrix factorization,'' \emph{Advances
  in Neural Information Processing Systems}, vol.~13, pp. 556--562, 2001.

\bibitem{Berry2007}
M.~W. Berry, M.~Browne, A.~N. Langville, V.~P. Pauca, and R.~J. Plemmons,
  ``Algorithms and applications for approximate nonnegative matrix
  factorization,'' \emph{Computational Statistics and Data Analysis}, vol.~52,
  no.~1, pp. 155--173, 2007.

\bibitem{Lin2007}
C.-J. Lin, ``Projected gradient methods for nonnegative matrix factorization,''
  \emph{Neural Computation}, vol.~19, no.~10, pp. 2756--2779, 2007.

\bibitem{Aharon2006}
M.~Aharon, M.~Elad, and A.~Bruckstein, ``{K-SVD}: An algorithm for designing
  overcomplete dictionaries for sparse representation,'' \emph{IEEE
  Transactions on Signal Processing}, vol.~54, no.~11, pp. 4311--4322, 2006.

\bibitem{Mairal2010}
J.~Mairal, F.~Bach, J.~Ponce, and G.~Sapiro, ``Online learning for matrix
  factorization and sparse coding,'' vol.~11, 2010.

\bibitem{Zhu2011Sparsity}
Zhu, Hao, Leus, Geert, Giannakis, Georgios, and B., ``Sparsity-cognizant total
  least-squares for perturbed compressive sampling.'' \emph{IEEE Transactions
  on Signal Processing}, vol.~59, no.~5, pp. 2002--2016, 2011.

\bibitem{Lin2010}
Z.~Lin, M.~Chen, and Y.~Ma, ``The augmented {Lagrange} multiplier method for
  exact recovery of corrupted low-rank matrices,''
  \emph{http://export.arxiv.org/abs/1009.5055}, 2010.

\bibitem{Wen2012}
Z.~Wen, W.~Yin, and Y.~Zhang, ``Solving a low-rank factorization model for
  matrix completion by a nonlinear successive over-relaxation algorithm,''
  \emph{Mathematical Programming Computation}, vol.~4, no.~4, pp. 333--361,
  2012.

\bibitem{Donoho2010a}
D.~L. Donoho, A.~Maleki, and A.~Montanari, ``Message passing algorithms for
  compressed sensing: {I}. motivation and construction,'' in \emph{2010 IEEE
  Information Theory Workshop on Information Theory (ITW 2010, Cairo)}, Jan
  2010, pp. 1--5.

\bibitem{Donoho2010b}
------, ``Message passing algorithms for compressed sensing: {II}. analysis and
  validation,'' in \emph{2010 IEEE Information Theory Workshop on Information
  Theory (ITW 2010, Cairo)}, Jan 2010, pp. 1--5.

\bibitem{Kschischang2001}
F.~Kschischang, B.~Frey, and H.-A. Loeliger, ``Factor graphs and the
  sum-product algorithm,'' \emph{{IEEE} Trans. Inform. Theory}, vol.~47, no.~2,
  pp. 498--519, Feb. 2001.

\bibitem{rangan2011generalized}
S.~Rangan, ``Generalized approximate message passing for estimation with random
  linear mixing,'' in \emph{Proc. Int. Symp. Inf. Theory}.\hskip 1em plus 0.5em
  minus 0.4em\relax IEEE, July. 2011, pp. 2168--2172.

\bibitem{Parker2014I}
J.~T. Parker, P.~Schniter, and V.~Cevher, ``Bilinear generalized approximate
  message passing {Part} ii: Applications,'' \emph{IEEE Transactions on Signal
  Processing}, vol.~62, no.~22, pp. 5854--5867, 2014.

\bibitem{rangan2019}
S.~{Rangan}, P.~{Schniter}, and A.~K. {Fletcher}, ``Vector approximate message
  passing,'' \emph{IEEE Transactions on Information Theory}, vol.~65, no.~10,
  pp. 6664--6684, Oct 2019.

\bibitem{Sarkar2019}
S.~{Sarkar}, A.~K. {Fletcher}, S.~{Rangan}, and P.~{Schniter}, ``Bilinear
  recovery using adaptive vector-amp,'' \emph{IEEE Transactions on Signal
  Processing}, vol.~67, no.~13, pp. 3383--3396, July 2019.

\bibitem{guo2015approximate}
Q.~Guo and J.~Xi, ``Approximate message passing with unitary transformation,''
  \emph{arXiv preprint arXiv:1504.04799}, Apr. 2015.

\bibitem{winn2005variational}
J.~Winn and C.~M. Bishop, ``Variational message passing,'' \emph{J. Mach.
  Learn. Res.}, vol.~6, pp. 661--694, Apr. 2005.

\bibitem{Glanz2013}
H.~Glanz and L.~Carvalho, ``An expectation-maximization algorithm for the
  matrix normal distribution,'' \emph{Statistics}, 2013.

\bibitem{Waal1985}
D.~J. De~Waal, \emph{Matrix-Valued Distributions}.\hskip 1em plus 0.5em minus
  0.4em\relax Encyclopedia of Statistical Sciences, 1985.

\bibitem{donoho2009message}
D.~L. Donoho, A.~Maleki, and A.~Montanari, ``Message-passing algorithms for
  compressed sensing,'' \emph{Proc. Nat. Acad. Sci.}, vol. 106, no.~45, pp.
  18\,914--18\,919, Nov. 2009.

\bibitem{rangan2019convergence}
S.~Rangan, P.~Schniter, A.~K. Fletcher, and S.~Sarkar, ``On the convergence of
  approximate message passing with arbitrary matrices,'' \emph{IEEE Trans. Inf.
  Theory}, vol.~65, no.~9, pp. 5339--5351, Sept. 2019.

\bibitem{guo2013}
Q.~Guo, D.~D. Huang, S.~Nordholm, J.~Xi, and Y.~Yu, ``Iterative frequency
  domain equalization with generalized approximate message passing,''
  \emph{IEEE Signal Process. Lett.}, vol.~20, no.~6, pp. 559--562, June. 2013.

\bibitem{jordan1999introduction}
M.~I. Jordan, Z.~Ghahramani, T.~S. Jaakkola, and L.~K. Saul, ``An introduction
  to variational methods for graphical models,'' \emph{Mach. Learn.}, vol.~37,
  no.~2, pp. 183--233, Nov. 1999.

\bibitem{Dauwels2007}
J.~Dauwels, ``On variational message passing on factor graphs,'' in \emph{IEEE
  International Symposium on Information Theory}, 2007, pp. 2546--2550.

\bibitem{matrixcook}
K.~B. Petersen and M.~S. Pedersen, \emph{The Matrix Cookbook}.\hskip 1em plus
  0.5em minus 0.4em\relax on-line, Sep. 2007, http://matrixcookbook.com.

\bibitem{Parker2014II}
J.~T. Parker, P.~Schniter, and V.~Cevher, ``Bilinear generalized approximate
  message passing {Part} ii: Applications,'' \emph{IEEE Transactions on Signal
  Processing}, vol.~62, no.~22, pp. 5854--5867, 2014.

\bibitem{Dias2012}
J.~M. Bioucas-Dias, A.~Plaza, N.~Dobigeon, M.~Parente, Q.~Du, P.~Gader, and
  J.~Chanussot, ``Hyperspectral unmixing overview: Geometrical, statistical,
  and sparse regression-based approaches,'' \emph{IEEE Journal of Selected
  Topics in Applied Earth Observations and Remote Sensing}, vol.~5, no.~2, pp.
  354--379, 2012.

\bibitem{Xu2003}
W.~Xu, ``Document clustering based on non-negative matrix factorization,''
  \emph{Proceedings of SIGIR-03}, 2003.

\bibitem{Brunet2004}
J.~P. Brunet, P.~Tamayo, T.~R. Golub, and J.~P. Mesirov, ``Metagenes and
  molecular pattern discovery using matrix factorization,'' \emph{Proceedings
  of the National Academy of Sciences}, vol. 101, no.~12, pp. 4164--4169, 2004.

\bibitem{Hoyer2002Non}
P.~O. Hoyer, ``Non-negative sparse coding,'' in \emph{Neural Networks for
  Signal Processing, 2002. Proceedings of the 2002 12th IEEE Workshop on},
  2002.

\end{thebibliography}


\begin{thebibliography}{10}
\providecommand{\url}[1]{#1}
\csname url@samestyle\endcsname
\providecommand{\newblock}{\relax}
\providecommand{\bibinfo}[2]{#2}
\providecommand{\BIBentrySTDinterwordspacing}{\spaceskip=0pt\relax}
\providecommand{\BIBentryALTinterwordstretchfactor}{4}
\providecommand{\BIBentryALTinterwordspacing}{\spaceskip=\fontdimen2\font plus
\BIBentryALTinterwordstretchfactor\fontdimen3\font minus
  \fontdimen4\font\relax}
\providecommand{\BIBforeignlanguage}[2]{{%
\expandafter\ifx\csname l@#1\endcsname\relax
\typeout{** WARNING: IEEEtran.bst: No hyphenation pattern has been}%
\typeout{** loaded for the language `#1'. Using the pattern for}%
\typeout{** the default language instead.}%
\else
\language=\csname l@#1\endcsname
\fi
#2}}
\providecommand{\BIBdecl}{\relax}
\BIBdecl

\bibitem{Boccardi2014}
F.~Boccardi, R.~W. Heath, A.~Lozano, T.~L. Marzetta, and P.~Popovski, ``Five
  disruptive technology directions for {5G},'' \emph{IEEE Communications
  Magazine}, vol.~52, no.~2, pp. 74--80, 2014.

\bibitem{Tullberg2016}
H.~Tullberg, P.~Popovski, Z.~Li, M.~A. Uusitalo, A.~Hoglund, O.~Bulakci,
  M.~Fallgren, and J.~F. Monserrat, ``The metis {5G} system concept: Meeting
  the {5G} requirements,'' \emph{IEEE Communications Magazine}, vol.~54,
  no.~12, pp. 132--139, 2016.

\bibitem{Islam2014}
M.~T. Islam, A.-e.~M. Taha, and S.~Akl, ``A survey of access management
  techniques in machine type communications,'' \emph{IEEE Communications
  Magazine}, vol.~52, no.~4, pp. 74--81, 2014.

\bibitem{Chen2018Sparse}
Z.~Chen, F.~Sohrabi, and W.~Yu, ``Sparse activity detection for massive
  connectivity,'' \emph{IEEE Transactions on Signal Processing}, vol.~66,
  no.~7, pp. 1890--1904, 2018.

\bibitem{Fu2019Active}
J.~Fu, G.~Wu, Y.~Zhang, L.~Deng, and S.~Fang, ``Active user identification
  based on asynchronous sparse bayesian learning with {SVM},'' \emph{IEEE
  Access}, vol.~7, pp. 108\,116--108\,124, 2019.

\bibitem{Takanori2022LWC}
T.~Hara and K.~Ishibashi, ``Blind multiple measurement vector amp based on
  expectation maximization for grant-free {NOMA},'' \emph{IEEE Wireless
  Communications Letters}, vol.~11, no.~6, pp. 1201--1205, 2022.

\bibitem{Zhang2020TVT}
X.~Zhang, F.~Labeau, L.~Hao, and J.~Liu, ``Joint active user detection and
  channel estimation via bayesian learning approaches in {MTC}
  communications,'' \emph{IEEE Transactions on Vehicular Technology}, vol.~70,
  no.~6, pp. 6222--6226, 2021.

\bibitem{Zhang2022TWC}
X.~Zhang, P.~Fan, J.~Liu, and L.~Hao, ``Bayesian learning-based multiuser
  detection for grant-free {NOMA} systems,'' \emph{IEEE Transactions on
  Wireless Communications}, vol.~21, no.~8, pp. 6317--6328, 2022.

\bibitem{Chen2022TWC}
W.~Chen, H.~Xiao, L.~Sun, and B.~Ai, ``Joint activity detection and channel
  estimation in massive {MIMO} systems with angular domain enhancement,''
  \emph{IEEE Transactions on Wireless Communications}, vol.~21, no.~5, pp.
  2999--3011, 2022.

\bibitem{Zhang2020IOT}
Z.~Zhang, Y.~Li, C.~Huang, Q.~Guo, L.~Liu, C.~Yuen, and Y.~L. Guan, ``User
  activity detection and channel estimation for grant-free random access in leo
  satellite-enabled internet of things,'' \emph{IEEE Internet of Things
  Journal}, vol.~7, no.~9, pp. 8811--8825, 2020.

\bibitem{Wei2017Approx}
C.~Wei, H.~Liu, Z.~Zhang, J.~Dang, and L.~Wu, ``Approximate message
  passing-based joint user activity and data detection for {NOMA},'' \emph{IEEE
  Communications Letters}, vol.~21, no.~3, pp. 640--643, 2017.

\bibitem{Chi2017Message}
Y.~Chi, L.~Liu, G.~Song, C.~Yuen, Y.~L. Guan, and Y.~Li, ``Message passing in
  c-ran: Joint user activity and signal detection,'' in \emph{GLOBECOM 2017 -
  2017 IEEE Global Communications Conference}, 2017, pp. 1--6.

\bibitem{Wei2019Message}
F.~Wei, W.~Chen, Y.~Wu, J.~Ma, and T.~A. Tsiftsis, ``Message-passing receiver
  design for joint channel estimation and data decoding in uplink grant-free
  scma systems,'' \emph{IEEE Transactions on Wireless Communications}, vol.~18,
  no.~1, pp. 167--181, 2019.

\bibitem{Du2018Joint}
Y.~Du, B.~Dong, W.~Zhu, P.~Gao, Z.~Chen, X.~Wang, and J.~Fang, ``Joint channel
  estimation and multiuser detection for uplink grant-free {NOMA},'' \emph{IEEE
  Wireless Communications Letters}, vol.~7, no.~4, pp. 682--685, 2018.

\bibitem{Liu2018Sparse}
L.~Liu, E.~G. Larsson, W.~Yu, P.~Popovski, C.~Stefanovic, and E.~de~Carvalho,
  ``Sparse signal processing for grant-free massive connectivity: A future
  paradigm for random access protocols in the internet of things,'' \emph{IEEE
  Signal Processing Magazine}, vol.~35, no.~5, pp. 88--99, 2018.

\bibitem{Zhang2018TVT}
Y.~Zhang, Q.~Guo, Z.~Wang, J.~Xi, and N.~Wu, ``Block sparse bayesian learning
  based joint user activity detection and channel estimation for grant-free
  {NOMA} systems,'' \emph{IEEE Transactions on Vehicular Technology}, vol.~67,
  no.~10, pp. 9631--9640, 2018.

\bibitem{Wang2015Compres}
B.~Wang, L.~Dai, Y.~Yuan, and Z.~Wang, ``Compressive sensing based multi-user
  detection for uplink grant-free non-orthogonal multiple access,'' pp. 1--5,
  2015.

\bibitem{Wang2016Dynamic}
B.~Wang, L.~Dai, Y.~Zhang, T.~Mir, and J.~Li, ``Dynamic compressive
  sensing-based multi-user detection for uplink grant-free {NOMA},'' \emph{IEEE
  Communications Letters}, vol.~20, no.~11, pp. 2320--2323, 2016.

\bibitem{Du2017Eff}
Y.~Du, B.~Dong, Z.~Chen, X.~Wang, Z.~Liu, P.~Gao, and S.~Li, ``Efficient
  multi-user detection for uplink grant-free {NOMA}: Prior-information aided
  adaptive compressive sensing perspective,'' \emph{IEEE Journal on Selected
  Areas in Communications}, vol.~35, no.~12, pp. 2812--2828, 2017.

\bibitem{Ding2019TWC}
T.~Ding, X.~Yuan, and S.~C. Liew, ``Sparsity learning-based multiuser detection
  in grant-free massive-device multiple access,'' \emph{IEEE Transactions on
  Wireless Communications}, vol.~18, no.~7, pp. 3569--3582, 2019.

\bibitem{Zhang2018TCOMM}
J.~Zhang, X.~Yuan, and Y.-J.~A. Zhang, ``Blind signal detection in massive
  {MIMO}: Exploiting the channel sparsity,'' \emph{IEEE Transactions on
  Communications}, vol.~66, no.~2, pp. 700--712, 2018.

\bibitem{Guo2017Mill}
Z.~Guo, X.~Wang, and W.~Heng, ``Millimeter-wave channel estimation based on 2-d
  beamspace music method,'' \emph{IEEE Transactions on Wireless
  Communications}, vol.~16, no.~8, pp. 5384--5394, 2017.

\bibitem{Xiao2017}
M.~Xiao, S.~Mumtaz, Y.~Huang, L.~Dai, Y.~Li, M.~Matthaiou, G.~K. Karagiannidis,
  E.~Bjornson, K.~Yang, C.-L. I, and A.~Ghosh, ``Millimeter wave communications
  for future mobile networks,'' \emph{IEEE Journal on Selected Areas in
  Communications}, vol.~35, no.~9, pp. 1909--1935, 2017.

\bibitem{Yuan2022UAMPMF}
Z.~Yuan, Q.~Guo, Y.~C. Eldar, and Y.~Li, ``Unitary approximate message passing
  for matrix factorization,'' \emph{ArXiv}, vol. abs/2208.00422, 2022.

\bibitem{Donoho2010message}
D.~L. Donoho, A.~Maleki, and A.~Montanari, ``Message passing algorithms for
  compressed sensing: I. motivation and construction,'' pp. 1--5, 2010.

\bibitem{rangan2011}
S.~Rangan, ``Generalized approximate message passing for estimation with random
  linear mixing,'' pp. 2168--2172, 2011.

\bibitem{Rangan2019convergence}
S.~Rangan, P.~Schniter, A.~K. Fletcher, and S.~Sarkar, ``On the convergence of
  approximate message passing with arbitrary matrices,'' \emph{IEEE
  Transactions on Information Theory}, vol.~65, no.~9, pp. 5339--5351, 2019.

\bibitem{guo2013}
Q.~Guo, D.~D. Huang, S.~Nordholm, J.~Xi, and Y.~Yu, ``Iterative frequency
  domain equalization with generalized approximate message passing,''
  \emph{IEEE Signal Process. Lett.}, vol.~20, no.~6, pp. 559--562, June. 2013.

\bibitem{guo2015approximate}
Q.~Guo and J.~Xi, ``Approximate message passing with unitary transformation,''
  \emph{arXiv preprint arXiv:1504.04799}, Apr. 2015.

\bibitem{Yuan2021}
Z.~Yuan, Q.~Guo, and M.~Luo, ``Approximate message passing with unitary
  transformation for robust bilinear recovery,'' \emph{IEEE Transactions on
  Signal Processing}, vol.~69, pp. 617--630, 2021.

\bibitem{jordan1999introduction}
M.~I. Jordan, Z.~Ghahramani, T.~S. Jaakkola, and L.~K. Saul, ``An introduction
  to variational methods for graphical models,'' \emph{Mach. Learn.}, vol.~37,
  no.~2, pp. 183--233, Nov. 1999.

\bibitem{tse_viswanath_2005}
D.~Tse and P.~Viswanath, \emph{Fundamentals of Wireless Communication}.\hskip
  1em plus 0.5em minus 0.4em\relax Cambridge University Press, 2005.

\bibitem{Baron2010}
D.~Baron, S.~Sarvotham, and R.~G. Baraniuk, ``Bayesian compressive sensing via
  belief propagation,'' \emph{IEEE Transactions on Signal Processing}, vol.~58,
  no.~1, pp. 269--280, 2010.

\bibitem{Vila2013}
J.~P. Vila and P.~Schniter, ``Expectation-maximization gaussian-mixture
  approximate message passing,'' \emph{IEEE Transactions on Signal Processing},
  vol.~61, no.~19, pp. 4658--4672, 2013.

\bibitem{Parker2014}
J.~T. Parker, P.~Schniter, and V.~Cevher, ``Bilinear generalized approximate
  message passing-part {I}: Derivation,'' \emph{IEEE Transactions on Signal
  Processing}, vol.~62, no.~22, pp. 5839--5853, 2014.

\end{thebibliography}

\end{document}